\documentclass[twocolumn,showpacs,aps,prd,superscriptaddress,floatfix]{revtex4}

\usepackage{graphicx}
\usepackage{dcolumn}
\usepackage{amsmath}
\usepackage{epsfig}
\usepackage{rotating}

\RequirePackage{xspace}

\usepackage{relsize}
\def\babar{\mbox{\slshape B\kern-0.1em{\smaller A}\kern-0.1em
    B\kern-0.1em{\smaller A\kern-0.2em R}}}

\def\epem       {\ensuremath{e^+e^-}\xspace}

\def\ellell     {\ensuremath{\ell^+ \ell^-}\xspace}

\def\qqbar {\ensuremath{q\overline q}\xspace}

\def\piz   {\ensuremath{\pi^0}\xspace}

\def\pip   {\ensuremath{\pi^+}\xspace}
\def\pim   {\ensuremath{\pi^-}\xspace}
\def\pipi  {\ensuremath{\pi^+\pi^-}\xspace}
\def\pipm  {\ensuremath{\pi^\pm}\xspace}
\def\pimp  {\ensuremath{\pi^\mp}\xspace}

\def\Kbar  {\kern 0.2em\overline{\kern -0.2em K}{}\xspace}

\def\Kz    {\ensuremath{K^0}\xspace}
\def\Kzb   {\ensuremath{\Kbar^0}\xspace}
\def\KzKzb {\ensuremath{\Kz \kern -0.16em \Kzb}\xspace}
\def\Kp    {\ensuremath{K^+}\xspace}
\def\Km    {\ensuremath{K^-}\xspace}
\def\Kpm   {\ensuremath{K^\pm}\xspace}
\def\Kmp   {\ensuremath{K^\mp}\xspace}
\def\KpKm  {\ensuremath{\Kp \kern -0.16em \Km}\xspace}

\def\Kstarz  {\ensuremath{K^{*0}}\xspace}

\def\Dbar    {\kern 0.2em\overline{\kern -0.2em D}{}\xspace}

\def\Dz      {\ensuremath{D^0}\xspace}
\def\Dzb     {\ensuremath{\Dbar^0}\xspace}
\def\DzDzb   {\ensuremath{\Dz {\kern -0.16em \Dzb}}\xspace}
\def\Dp      {\ensuremath{D^+}\xspace}
\def\Dm      {\ensuremath{D^-}\xspace}

\def\DpDm    {\ensuremath{\Dp {\kern -0.16em \Dm}}\xspace}

\def\B       {\ensuremath{B}\xspace}
\def\Bbar    {\kern 0.18em\overline{\kern -0.18em B}{}\xspace}
\def\Bb      {\ensuremath{\Bbar}\xspace}
\def\BB      {\ensuremath{B\Bbar}\xspace} 
\def\Bz      {\ensuremath{B^0}\xspace}
\def\Bzb     {\ensuremath{\Bbar^0}\xspace}
\def\BzBzb   {\ensuremath{\Bz {\kern -0.16em \Bzb}}\xspace}
\def\Bu      {\ensuremath{B^+}\xspace}
\def\Bub     {\ensuremath{B^-}\xspace}
\def\Bp      {\ensuremath{\Bu}\xspace}
\def\Bm      {\ensuremath{\Bub}\xspace}
\def\Bpm     {\ensuremath{B^\pm}\xspace}

\def\BpBm    {\ensuremath{\Bu {\kern -0.16em \Bub}}\xspace}

\def\BorBbar    {\kern 0.18em\optbar{\kern -0.18em B}{}\xspace}
\def\DorDbar    {\kern 0.18em\optbar{\kern -0.18em D}{}\xspace}
\def\KorKbar    {\kern 0.18em\optbar{\kern -0.18em K}{}\xspace}

\def\jpsi     {\ensuremath{{J\mskip -3mu/\mskip -2mu\psi\mskip 2mu}}\xspace}
\def\psitwos  {\ensuremath{\psi{(2S)}}\xspace}

\def\chiczero {\ensuremath{\chi_{c0}}\xspace}

\mathchardef\Upsilon="7107
\def\Y#1S{\ensuremath{\Upsilon{(#1S)}}\xspace}

\def\FourS {\Y4S}

\mathchardef\Deltares="7101
\mathchardef\Xi="7104
\mathchardef\Lambda="7103
\mathchardef\Sigma="7106
\mathchardef\Omega="710A

\def\Deltabar{\kern 0.25em\overline{\kern -0.25em \Deltares}{}\xspace}
\def\Lbar{\kern 0.2em\overline{\kern -0.2em\Lambda\kern 0.05em}\kern-0.05em{}\xspace}
\def\Sigbar{\kern 0.2em\overline{\kern -0.2em \Sigma}{}\xspace}
\def\Xibar{\kern 0.2em\overline{\kern -0.2em \Xi}{}\xspace}
\def\Obar{\kern 0.2em\overline{\kern -0.2em \Omega}{}\xspace}
\def\Nbar{\kern 0.2em\overline{\kern -0.2em N}{}\xspace}
\def\Xb{\kern 0.2em\overline{\kern -0.2em X}{}\xspace}

\def\BR         {{\ensuremath{\cal B}\xspace}}

\def\mes        {\mbox{$m_{\rm ES}$}\xspace}

\def\DeltaE     {\mbox{$\Delta E$}\xspace}

\newcommand{\tev}{\ensuremath{\mathrm{\,Te\kern -0.1em V}}\xspace}
\newcommand{\gev}{\ensuremath{\mathrm{\,Ge\kern -0.1em V}}\xspace}
\newcommand{\mev}{\ensuremath{\mathrm{\,Me\kern -0.1em V}}\xspace}
\newcommand{\kev}{\ensuremath{\mathrm{\,ke\kern -0.1em V}}\xspace}
\newcommand{\ev}{\ensuremath{\mathrm{\,e\kern -0.1em V}}\xspace}
\newcommand{\gevc}{\ensuremath{{\mathrm{\,Ge\kern -0.1em V\!/}c}}\xspace}
\newcommand{\mevc}{\ensuremath{{\mathrm{\,Me\kern -0.1em V\!/}c}}\xspace}
\newcommand{\gevcc}{\ensuremath{{\mathrm{\,Ge\kern -0.1em V\!/}c^2}}\xspace}
\newcommand{\mevcc}{\ensuremath{{\mathrm{\,Me\kern -0.1em V\!/}c^2}}\xspace}

\def\invfb   {\ensuremath{\mbox{\,fb}^{-1}}\xspace}

\def\mus  {\ensuremath{\rm \,\mus}\xspace}

\def\mus        {\ensuremath{\,\mu{\rm s}}\xspace}    

\def\to                 {\ensuremath{\rightarrow}\xspace}

\def\pep2{PEP-II}

\newcommand{\chisq}{\ensuremath{\chi^2}\xspace}

\def\gsim{{~\raise.15em\hbox{$>$}\kern-.85em
          \lower.35em\hbox{$\sim$}~}\xspace}
\def\lsim{{~\raise.15em\hbox{$<$}\kern-.85em
          \lower.35em\hbox{$\sim$}~}\xspace}

\def\CP                {\ensuremath{C\!P}\xspace}

\xspace

\def\jetset74   {\mbox{\tt Jetset \hspace{-0.5em}7.\hspace{-0.2em}4}\xspace}
\renewcommand{\eqref}[1]{Eq.~(\ref{eq:#1})}

\newcommand{\splot}    {\mbox{$_s{\cal P}lot$}\xspace}

\newcommand{\onreslumi}  {\mbox{347.5\invfb}}
\newcommand{\offreslumi} {\mbox{36.6\invfb}}
\newcommand{\bbpairs}    {\mbox{$(383.2\pm4.2)\times10^{6}$}}

\newcommand{\nbb}        {\mbox{$N_{\BB}$}}

\newcommand{\Lzero}      {\mbox{$L_0$}}
\newcommand{\Ltwo}       {\mbox{$L_2$}}

\newcommand{\sigmadeltae} {\mbox{$\sigma_{\Delta E}$}}
\newcommand{\DeltaEsig}   {\mbox{$\frac{\DeltaE}{\sigmadeltae}$}}
\newcommand{\DeltaEpr}    {\mbox{$\DeltaE^{\prime}$}}
\newcommand{\mACSq}       {\mbox{$m^2_{K\pi}$}}
\newcommand{\mBCSq}       {\mbox{$m^2_{\pi\pi}$}}

\newcommand{\Kpppos}             {\mbox{$\Kp   \pim  \pip$}}

\newcommand{\BtoKpppos}          {\mbox{$\Bp \to \Kpppos$}}

\newcommand{\KPP}                {\mbox{$\Kpm  \pimp \pipm$}}
\newcommand{\KKP}                {\mbox{$\Kpm  \Kmp  \pipm$}}

\newcommand{\BtoKPP}             {\mbox{$\Bpm \to \KPP$}}
\newcommand{\BtoKKP}             {\mbox{$\Bpm \to \KKP$}}

\def\Kpi   {\ensuremath{K^+\pi^-}\xspace}

\newcommand{\KstarI}             {\mbox{$\Kstarz(892)$}}
\newcommand{\KstarIpip}          {\mbox{$\KstarI \pip$}}
\newcommand{\BptoKstarIpip}      {\mbox{$\Bp \to \KstarIpip$}}
\newcommand{\KstarItoKppim}      {\mbox{$\KstarI \to \Kp \pim$}}

\newcommand{\KstarII}            {\mbox{$\Kstarz_{0}(1430)$}}
\newcommand{\KstarIIpip}         {\mbox{$\KstarII \pip$}}
\newcommand{\BptoKstarIIpip}     {\mbox{$\Bp \to \KstarIIpip$}}

\newcommand{\KpiSwave}           {\mbox{$(K\pi)^{*0}_0$}}
\newcommand{\KpiSwavepip}        {\mbox{$\KpiSwave \pip$}}
\newcommand{\KpiSwavetoKppim}    {\mbox{$\KpiSwave \to \Kp \pim$}}

\newcommand{\KstarIII}           {\mbox{$\Kstarz_{2}(1430)$}}
\newcommand{\KstarIIIpip}        {\mbox{$\KstarIII \pip$}}
\newcommand{\BptoKstarIIIpip}    {\mbox{$\Bp \to \KstarIIIpip$}}
\newcommand{\KstarIIItoKppim}    {\mbox{$\KstarIII \to \Kp \pim$}}

\newcommand{\KstarIV}            {\mbox{$\Kstarz(1680)$}}

\newcommand{\rhoz}               {\mbox{$\rho^0$}}

\newcommand{\rhoI}               {\mbox{$\rhoz(770)$}}
\newcommand{\rhoIKpm}            {\mbox{$\rhoI \Kpm$}}
\newcommand{\rhoIKp}             {\mbox{$\rhoI \Kp$}}

\newcommand{\BtorhoIK}           {\mbox{$\Bpm \to \rhoIKpm$}}
\newcommand{\BptorhoIKp}         {\mbox{$\Bp \to \rhoIKp$}}

\newcommand{\rhoItopippim}       {\mbox{$\rhoI \to \pip\pim$}}

\newcommand{\omegaI}              {\mbox{$\omega(782)$}}
\newcommand{\omegaIKp}            {\mbox{$\omegaI \Kp$}}

\newcommand{\BptoomegaIKp}        {\mbox{$\Bp \to \omegaIKp$}}

\newcommand{\omegaItopippim}      {\mbox{$\omegaI \to \pip\pim$}}

\newcommand{\fz}                 {\mbox{$f_0$}}
\newcommand{\fI}                 {\mbox{$\fz(980)$}}
\newcommand{\fIKp}               {\mbox{$\fI \Kp$}}

\newcommand{\fItopippim}         {\mbox{$\fI \to \pip\pim$}}

\newcommand{\rhoII}              {\mbox{$\rhoz(1450)$}}
\newcommand{\rhoIIKp}             {\mbox{$\rhoII \Kp$}}

\newcommand{\BptorhoIIKp}         {\mbox{$\Bp \to \rhoIIKp$}}

\newcommand{\rhoIItopippim}       {\mbox{$\rhoII \to \pip\pim$}}

\newcommand{\fII}                {\mbox{$f_2(1270)$}}
\newcommand{\fIIKp}              {\mbox{$\fII \Kp$}}

\newcommand{\BptofIIKp}          {\mbox{$\Bp \to \fIIKp$}}

\newcommand{\fIItopippim}        {\mbox{$\fII \to \pip\pim$}}

\newcommand{\fIV}                {\mbox{$f_0(1500)$}}

\newcommand{\fVI}                 {\mbox{$f_{\rm X}(1300)$}}
\newcommand{\fVIKp}               {\mbox{$\fVI \Kp$}}

\newcommand{\fVItopippim}         {\mbox{$\fVI \to \pip\pim$}}

\newcommand{\chiczKp}            {\mbox{$\chiczero \Kp$}}

\newcommand{\chicztopippim}      {\mbox{$\chiczero \to \pip\pim$}}

\newcommand{\Dzbpip}             {\mbox{$\Dzb \pip$}}
\newcommand{\BptoDzbpip}         {\mbox{$\Bp \to \Dzbpip$}}

\newcommand{\DzbtoKpi}           {\mbox{$\Dzb \to \Kp\pim$}}

\newcommand{\JPsitoll}           {\mbox{$\jpsi \to \ellell$}}

\newcommand{\Psitoll}            {\mbox{$\psitwos \to \ellell$}}

\newcommand{\NonRes}             {\mbox{\Kpppos\ nonresonant}}

\def\ACP {{\ensuremath{A_{\CP}}\xspace}}

\newcommand{\fcc}[1]{\multicolumn{4}{c}{#1}}

\newcommand{\Dx}{\mbox{$\Delta x$}}
\newcommand{\Dy}{\mbox{$\Delta y$}}

\newcommand{\shortfigref}[1]{Fig.~\ref{fig:#1}}

\newcommand{\BABARPubYear}    {08}
\newcommand{\BABARPubNumber}  {005}

\newcommand{\SLACPubNumber} {13189}
\newcommand{\LANLNumber} {0803.4451}

\def\figurebox#1#2#3{%
    \def\arg{#3}%
    \ifx\arg\empty
    {\hfill\vbox{\hsize#2\hrule\hbox to #2{\vrule\hfill\vbox to #1{\hsize#2\vfill}\vrule}\hrule}\hfill}%
    \else
    {\hfill\epsfbox{#3}\hfill}%
    \fi}

\begin{document}


\begin{flushleft}
arXiv:\LANLNumber\ [hep-ex] \\
SLAC-PUB-\SLACPubNumber \\
\babar-PUB-\BABARPubYear/\BABARPubNumber
\end{flushleft}

\title{
  {
    \large \bf \boldmath 
    Evidence for Direct \CP\ Violation from 
    Dalitz-plot analysis of \BtoKPP
  }
}

%
\author{B.~Aubert}
\author{M.~Bona}
\author{Y.~Karyotakis}
\author{J.~P.~Lees}
\author{V.~Poireau}
\author{E.~Prencipe}
\author{X.~Prudent}
\author{V.~Tisserand}
\affiliation{Laboratoire de Physique des Particules, IN2P3/CNRS et Universit\'e de Savoie, F-74941 Annecy-Le-Vieux, France }
\author{J.~Garra~Tico}
\author{E.~Grauges}
\affiliation{Universitat de Barcelona, Facultat de Fisica, Departament ECM, E-08028 Barcelona, Spain }
% ...
\collaboration{The \babar\ Collaboration}
\noaffiliation

\date{
  \today
}

\begin{abstract} 
\noindent

We report a Dalitz-plot analysis of the charmless hadronic decays of charged
\B\ mesons to the final state \KPP.  
Using a sample of \bbpairs\ \BB\ pairs collected by the \babar\ detector, 
we measure \CP-averaged branching fractions and direct \CP\ asymmetries
for intermediate resonant and nonresonant contributions.
We find evidence for direct \CP\ violation in the decay \BtorhoIK,
with a \CP-violation parameter $A_{\CP} = (+44\pm10\pm4\,^{+5}_{-13})\%$ .

\end{abstract}

\pacs{13.25.Hw, 12.15.Hh, 11.30.Er}
\maketitle

\section{Introduction}

The Kobayashi-Maskawa mechanism of \CP\ nonconservation~\cite{Kobayashi:1973fv}
has, in recent years, been confirmed through 
observations of direct \CP\ violation in the kaon
system~\cite{AlaviHarati:1999xp,Fanti:1999nm}
and of both mixing-induced~\cite{Aubert:2001nu,Abe:2001xe}
and direct~\cite{Aubert:2004qm} \CP\ violation in the $B$ meson system.
However, it is striking that despite an enormous experimental effort~\cite{pdg2006,Barberio:2007cr},
no observation of \CP\ violation in the decay of any charged particle
has yet been made.
In some cases, for example, in $\Bpm \to \pi^0 \Kpm$~\cite{Aubert:2007hh,Lin:2008zz},
the absence of an asymmetry is difficult to understand theoretically~\cite{Buras:2003dj,Buras:2005cv,Baek:2004rp,Kim:2005jp,Li:2005wx,Li:2005kt}.
The search for such effects is therefore a priority for studies 
of the weak interaction.

Several theoretical investigations have suggested that large \CP\ asymmetries
may be observed in $\BtorhoIK$
decays~\cite{Chen:2001jx,Du:2002up,Beneke:2003zv,Chiang:2003pm,Isola:2003fh,Li:2006jb,Li:2006jv,ElBennich:2006yi,Wang:2008rk}.
Because of the large width of the $\rho^0$ meson, 
this channel must be studied using a Dalitz-plot analysis of the
$\KPP$ final state.
Previous measurements found 
$\ACP(\BtorhoIK) = (+32 \pm 13 \pm 6 \,^{+8}_{-5})\%$~\cite{Aubert:2005ce} and
$\ACP(\BtorhoIK) = (+30 \pm 11 \pm 2 \,^{+11}_{-4})\%$~\cite{Garmash:2005rv}
(where the errors are statistical, systematic, and model-dependent, respectively),
indicating evidence for \CP\ violation in this decay,
and leaving a strong need for more precise and conclusive studies.

Several other considerations motivate a precise analysis of \BtoKPP\ decays.
The \CP\ asymmetries in the decays 
$\Bp \to \KstarI \pip$, $\Bp \to \KstarII \pip$, and $\Bp \to \KstarIII \pip$
are predicted to be negligible~\cite{Beneke:2003zv,Chiang:2003pm}
compared to the current precision,
since these are mediated by $b \to s$ loop (penguin) transitions only,
with no $b \to u$ tree component.
A significant \CP-violation effect in any of these channels would 
therefore provide a signature of new physics.
On the other hand, decays such as $\Bp \to \rho^0 \Kp$ may have comparable
contributions from both tree and penguin amplitudes,
and their interference
is sensitive to their relative weak (\CP\ violating) phase difference $\gamma$.
Results from amplitude analyses of \BtoKPP\ and other 
$B \to K\pi\pi$ decays~\cite{Garmash:2006fh,Aubert:2007vi,Aubert:2007bs}
can be combined to constrain the hadronic parameters
and allow a relatively clean determination of $\gamma$~\cite{Lipkin:1991st,Deshpande:2002be,Ciuchini:2006kv,Gronau:2006qn,Gronau:2007vr,Bediaga:2007zz}.

Understanding the dynamics of the \BtoKPP\ Dalitz plot is also a priority.
Previous studies have left unresolved the nature of a structure 
peaking at invariant mass around $1300 \ \mevcc$ in the $\pip\pim$ spectrum
(denoted \fVI\ in this work).
Similar enhancements have also been seen at low
$\Kp\Km$ invariant masses in $\B \to KKK$~\cite{Garmash:2004wa,Aubert:2006nu,Aubert:2007sd}
and \BtoKKP~\cite{Aubert:2007xb} decays.
The origin of these structures
has aroused considerable interest among theorists~\cite{Minkowski:2004xf,Furman:2005xp,Cheng:2005nb,Wang:2006ri,Cheng:2007si},
as it is of great importance in the understanding of low energy spectroscopy~\cite{Klempt:2007cp}.

In this paper we present results from an amplitude analysis of
\BtoKPP\ decays based on a \onreslumi\ data sample containing
\bbpairs\ \BB\ pairs (\nbb).
Compared to our previous publication~\cite{Aubert:2005ce}, 
we have increased the data sample by $70\%$,
included several improvements in reconstruction algorithms 
that enhance the signal efficiency,
made several modifications to the analysis to increase the sensitivity
to direct \CP-violating effects (for example, 
by including more discriminating variables in the maximum likelihood fit),
and improved our model of the Dalitz-plot structure.
Moreover, we have developed a novel parametrization of the coefficients used 
in the fit that ensures good statistical behavior of the fitted parameters.

The data were collected with the \babar\ detector~\cite{babar} at the
SLAC \pep2\ asymmetric-energy \epem\ storage rings~\cite{pep2} operating at
the $\FourS$ resonance with center-of-mass (CM) energy of $\sqrt{s}=10.58\gev$.
An additional total integrated luminosity of \offreslumi\ was recorded
$40\mev$ below the $\FourS$ resonance (``off-peak'' data)  and was used to
study backgrounds from continuum production.

\section{Amplitude Analysis Formalism}

A number of intermediate states contribute to the decay \BtoKPP.
Their individual contributions are obtained from a maximum likelihood
fit of the distribution of events in the Dalitz plot formed from the two
variables $\mACSq \equiv m_{\Kpm\pimp}^2$ and $\mBCSq \equiv
m_{\pipm\pimp}^2$.
The total signal amplitudes for \Bp\ and \Bm\ decays are given, in the
isobar formalism (see for example~\cite{isobar1,isobar2,isobar3}), by:
\begin{eqnarray}
A = A(\mACSq,\mBCSq) &=& \sum_j c_j F_j(\mACSq,\mBCSq) \\
\overline{A} = \overline{A}(\mACSq,\mBCSq) &=& \sum_j \overline{c}_j \overline{F}_j(\mACSq,\mBCSq) ,.
\end{eqnarray}
The complex coefficient for a given decay mode $j$ is $c_j$ and is measured
relative to one of the contributing channels (\KstarI\ in this analysis).
The $c_j$ contain all the weak phase dependence and so $F_j \equiv \overline{F}_j$.
The distributions $F_j$ describe the dynamics of the decay amplitudes and
are the product of an invariant mass term ($R_j$), two Blatt-Weisskopf
barrier form factors ($X_J(z)$)~\cite{blatt-weisskopf}, and an angular
function ($T_j$):
\begin{equation}
F_j = R_j \times X_J(p^{\star}) \times X_J(q) \times T_j\,,
\end{equation}
where $J$ is the spin of the resonance, $q$ is the momentum of either
daughter in the rest frame of the resonance, and $p^{\star}$ is the momentum
of the bachelor particle in the rest frame of the \B.
The $F_j$ are normalized over the entire Dalitz plot:
\begin{equation}
\int\!\!\int \left|F_j(\mACSq,\mBCSq)\right|^2 d\mACSq d\mBCSq = 1\,.
\end{equation}

The Blatt-Weisskopf barrier form factors are given by:
\begin{eqnarray}
X_{J=0}(z) & = & 1, \\
X_{J=1}(z) & = & \sqrt{1/(1 + (z \, r_{\rm BW})^2)}, \\
X_{J=2}(z) & = & \sqrt{1/((z \, r_{\rm BW})^4 + 3(z \, r_{\rm BW})^2 + 9)},
\label{eqn:BlattEqn}
\end{eqnarray}
where $r_{\rm BW}$, the meson radius parameter, is taken to be
$(4.0\pm1.0)\,(\!\gevc)^{-1}$~\cite{buggpc}.

For most resonances in this analysis the $R_j$ are taken to be relativistic
Breit-Wigner line shapes:
\begin{equation}
R_j(m) = \frac{1}{(m^2_0 - m^2) - i m_0 \Gamma(m)},
\label{eqn:BreitWigner}
\end{equation}
where $m_0$ is the nominal mass of the resonance and $\Gamma(m)$ is the
mass-dependent width.
In the general case of a spin-$J$ resonance, the latter can be expressed as
\begin{equation}
\Gamma(m) = \Gamma_0 \left( \frac{q}{q_0}\right)^{2J+1} 
\left(\frac{m_0}{m}\right) \frac{X^2_J(q)}{X^2_J(q_0)}.
\label{eqn:resWidth}
\end{equation}
The symbol $\Gamma_0$ denotes the nominal width of the resonance.
The values of $m_0$ and $\Gamma_0$ are obtained from standard
tables~\cite{pdg2006}.
The symbol $q_0$ denotes the value of $q$ when $m = m_0$.

For the \fI\ line shape the Flatt\'e form~\cite{Flatte} is used.
In this case the mass-dependent width is given by the sum 
of the widths in the $\pi\pi$ and $KK$ systems:
\begin{equation}
\Gamma(m) = \Gamma_{\pi\pi}(m) + \Gamma_{KK}(m),
\label{eqn:FlatteW1}
\end{equation}
where
\begin{eqnarray}
\Gamma_{\pi\pi}(m) &=&
g_{\pi} \Bigg(\frac{1}{3}\sqrt{1 - 4m_{\piz}^2/m^2} + \\ \nonumber
&& \phantom{g_{\pi} \Bigg(\frac{1}{3}} \frac{2}{3}\sqrt{1 - 4m_{\pipm}^2/m^2}\Bigg)\,,\\
\Gamma_{KK}(m) &=&
g_{K} \Bigg(\frac{1}{2}\sqrt{1 - 4m_{\Kpm}^2/m^2} + \\ \nonumber
&& \phantom{g_{K} \Bigg(\frac{1}{2}} \frac{1}{2}\sqrt{1 - 4m_{\Kz}^2/m^2}\Bigg)\,.
\label{eqn:FlatteW2}
\end{eqnarray}
The fractional coefficients arise from isospin conservation and $g_{\pi}$
and $g_{K}$ are coupling constants for which we take the values:
\begin{eqnarray}
\label{eq:flatte-params}
g_{\pi} &=& \left(0.165\pm0.010\pm0.015\right)\gevcc\,,\\
\nonumber
g_K &=& \left(4.21\pm0.25\pm0.21\right) \times g_{\pi}\,,
\end{eqnarray}
from results obtained by the BES experiment~\cite{new-BES}.

The $0^+$ component of the $K\pi$ spectrum is not well
understood~\cite{LASS,bugg}; we dub this component \KpiSwave\ and
use the LASS parametrization~\cite{LASS} which consists of the
\KstarII\ resonance together with an effective range nonresonant component:
\begin{eqnarray}
\label{eq:LASSEqn}
R_j(m)  &=& \frac{m_{K\pi}}{q \cot{\delta_B} - iq} \\ &+& e^{2i \delta_B} 
\frac{m_0 \Gamma_0 \frac{m_0}{q_0}}
     {(m_0^2 - m_{K\pi}^2) - i m_0 \Gamma_0 \frac{q}{m_{K\pi}} \frac{m_0}{q_0}},
\nonumber
\end{eqnarray}
where $\cot{\delta_B} = \frac{1}{aq} + \frac{1}{2} r q$.
We have used the following values
for the scattering length and effective range parameters of this
distribution~\cite{Aubert:2005ce}:
\begin{eqnarray}
\label{eq:lass-params}
a &=& \left(2.07\pm0.10\right)(\!\gevc)^{-1} \\
\nonumber
r &=& \left(3.32\pm0.34\right)(\!\gevc)^{-1}\,,
\end{eqnarray}
and the effective range part of the amplitude is cut off at $m_{K\pi} =
1.8\gevcc$.
With our final fit model we have determined the preferred values of the $a$
and $r$ parameters and obtain results consistent with those given above.
Integrating separately the resonant part, the effective range part, and the
coherent sum we find that the \KstarII\ resonance accounts for 81\%, the
effective range term 45\%, and destructive interference between the two terms 
is responsible for the excess 26\%.

The nonresonant component of the Dalitz plot is modeled with a constant complex
amplitude.
We use alternative models~\cite{Garmash:2005rv} to evaluate the model
dependence of our results.
In these studies we also use the Gounaris-Sakurai form~\cite{GS} as an 
alternative model for the \rhoI.

For the angular distribution terms $T_j$ we follow the Zemach tensor
formalism~\cite{Zemach1,Zemach2}.  For the decay of a spin 0 \B\ meson into
a spin $J$ resonance and a spin 0 bachelor particle this
gives~\cite{asner-review}:
\begin{eqnarray}
\nonumber
T_j^{J=0} &=& 1 \\
\label{eq:angular}
T_j^{J=1} &=& -2 \, \vec{p}\cdot\vec{q} \\
T_j^{J=2} &=& \frac{4}{3} \left[3(\vec{p}\cdot\vec{q})^2 - (|\vec{p}\,||\vec{q}\,|)^2\right],
\nonumber
\end{eqnarray}
where $\vec{p}$ is the momentum of the bachelor particle and $\vec{q}$ is the
momentum of the resonance daughter with charge opposite from that of the
bachelor particle, both measured in the rest frame of the resonance.

The complex coefficients $c_j$ and $\overline{c}_j$ can be parametrized 
in various ways that take the possibility of direct \CP\ violation into
account.
We have investigated the choices used in previous studies~\cite{Asner:2003uz,Garmash:2005rv}
and found that they are susceptible to biases on the fitted parameters,
particularly when a resonant contribution is small in magnitude.
To ensure the good statistical behavior of our fit,
we parametrize the complex coefficients in the following way:
\begin{eqnarray}
c_j &=& (x_j + \Delta x_j) + i (y_j + \Delta y_j) \\ \nonumber
\overline{c}_j &=& (x_j - \Delta x_j) + i (y_j - \Delta y_j) \, .
\end{eqnarray}
We have verified that these parameters have approximately Gaussian behavior
independent of their true values.
In this approach, $\Delta x_j$ and $\Delta y_j$ are \CP-violating parameters.

To allow comparison among experiments we present also fit fractions ($\it FF$),
defined as the integral of a single decay amplitude squared divided by 
the coherent matrix element squared for the complete Dalitz plot:
\begin{equation}
{\it FF}_j =
\frac
{\displaystyle\int\!\!\int{\left(\left|c_j F_j\right|^2 + \left|\overline{c}_j \overline{F}_j\right|^2\right)} d\mACSq \, d\mBCSq}
{\displaystyle\int\!\!\int{\left(\left|A\right|^2 + \left|\overline{A}\right|^2\right)} d\mACSq \, d\mBCSq} \,.
\label{eq:fitfraction}
\end{equation}
The sum of all the fit fractions is not necessarily unity due to the potential
presence of net constructive or destructive interference.
The \CP\ asymmetry for a given intermediate state is easily determined
from the fitted parameters
\begin{eqnarray}
  \ACP_{\!,\,j} & = &
  \frac
  {\left|\overline{c}_j\right|^2 - \left|c_j\right|^2}
  {\left|\overline{c}_j\right|^2 + \left|c_j\right|^2} \nonumber \\
  & = & 
  \frac
  {-2\left(x_j \Delta x_j + y_j \Delta y_j\right)}
  {x_j^2 + \Delta x_j^2 + y_j^2 + \Delta y_j^2} \, .
  \label{eq:cpasym}
\end{eqnarray}

The signal Dalitz-plot probability density function (PDF) is formed from
the total amplitude as follows:
\begin{eqnarray}
  \label{eq:SigDPLikeEqn}
  &&
  {\cal L}_{\rm sig}(\mACSq,\mBCSq,q_{\B}) = \\ \nonumber
  &&
  \phantom{{\cal L}_{\rm sig}(\mACSq}
  \frac
  {
    \frac{1+q_{\B}}{2} |A|^2 \; \varepsilon + \frac{1-q_{\B}}{2} |\overline{A}|^2 \; \overline{\varepsilon}
  }
  {
    \displaystyle{\int\!\!\int \!\! \left(\, |A|^2 \, \varepsilon +
    |\overline{A}|^2 \, \overline{\varepsilon}\,\right) d\mACSq\,d\mBCSq}
  } \,,
\end{eqnarray}
where $q_{\B}$ is the charge of the \B\ meson candidate and $\varepsilon
\equiv \varepsilon(\mACSq,\mBCSq)$ and $\overline{\varepsilon} \equiv
\overline{\varepsilon}(\mACSq,\mBCSq)$ are the signal reconstruction
efficiencies for \Bp\ and \Bm\ events, respectively, defined for all points in
the Dalitz plot.

\section{Candidate Selection}

The \B\ candidates are reconstructed from events that have four or more
charged tracks.  Each track is required to be well measured and to
originate from the beam spot.  The \B\ candidates are formed from
three-charged-track combinations and particle identification criteria are
applied to reject electrons and to separate kaons and pions.
In our final state, the average selection efficiency for kaons that have
passed the tracking requirements is about 80\% including geometrical
acceptance, while the average misidentification probability of pions as
kaons is about 2\%.

\begin{figure}[htb]
\begin{center}
\includegraphics[angle=0, width=\columnwidth]{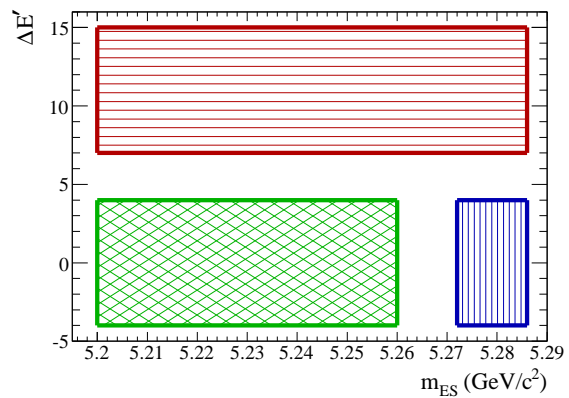}
\caption
{
Regions of the \DeltaEpr--\mes\ plane.
The red/horizontal hatching is the ``upper sideband,''
the green/crossed hatching is the ``left sideband,''
and the blue/vertical hatching is the ``signal region.''
}
\label{fig:regions}
\end{center}
\end{figure}

Two kinematic variables are used to identify signal \B\ decays.
The first variable is
\begin{equation}
\DeltaE = E_B^* - \sqrt{s}/2,
\end{equation}
the difference between the reconstructed CM energy of the \B-meson
candidate and $\sqrt{s}/2$, where $\sqrt{s}$ is the total CM energy.
The second is the energy-substituted mass
\begin{equation}
\mes = \sqrt{(s/2 + \vec{p}_i \cdot \vec{p}_\B )^2/ E_i^2 - |\vec{p}_\B|^2},
\end{equation}
where $\vec{p}_\B$ is the \B\ momentum and ($E_i$,$\vec{p}_i$) is the
four momentum of the initial state, all measured in the laboratory frame.
The \mes\ distribution for signal events peaks near the \B\ mass with a
resolution of around $2.4\mevcc$, while the \DeltaE\ distribution peaks
near zero with a resolution of around $19\mev$.
The resolution of \DeltaE\ is strongly dependent on the position in the
Dalitz plot and so instead of \DeltaE\ we use
\begin{equation}
\DeltaEpr = \DeltaEsig \,,
\end{equation}
where \sigmadeltae\ is the error on \DeltaE, determined separately
for each event.  This variable exhibits no such dependence.
We initially require events to lie in the region formed by the following
selection criteria: $5.200<\mes<5.286\gevcc$ and $-4.0<\DeltaEpr<15.0$.
The region of \DeltaEpr\ below $-4$ is heavily contaminated by four-body
\B\ backgrounds and so is not useful for studying the continuum background.
The selected region is then subdivided into three areas:
the ``left sideband'' region ($5.20<\mes<5.26\gevcc$ and
$-4.0<\DeltaEpr<4.0$) used to study the background \DeltaEpr\ and
Dalitz-plot distributions; the ``upper sideband'' region ($5.200<\mes<5.286\gevcc$
and $7.0<\DeltaEpr<15.0$) used to study the background \mes\ distributions;
and the ``signal region'' ($5.272<\mes<5.286\gevcc$ and
$-4.0<\DeltaEpr<4.0$) where the final fit to data is performed.
These three regions are illustrated in \shortfigref{regions}.
Following the calculation of these kinematic variables, each of the
\B\ candidates is refitted with its mass constrained to the world average
value of the \B-meson mass~\cite{pdg2006} in order to improve the
Dalitz-plot position resolution.

The dominant source of background comes from light quark and charm
continuum production ($\epem\to\qqbar$, where $q = u,d,s,c$).  This
background is suppressed by requirements on event-shape variables
calculated in the CM frame.
We compute a neural network (NN) from the following five variables: the
ratio of the Legendre polynomial moments \Lzero\ and \Ltwo~\cite{legendre},
the absolute value of the cosine of the angle between the direction of the
\B\ and the detector axis, the absolute value of the cosine of the angle
between the \B\ thrust axis and the detector axis, the output of a
multivariate \B-flavor tagging algorithm~\cite{tagging} multiplied by the
charge of the \B\ candidate, and the significance of the measured proper
time difference of the \B\ decay vertices.
The NN is trained using samples of off-resonance data and signal Monte
Carlo (MC) simulated events.
The selection requirement placed on the NN output, optimized with MC
events, accepts 61\% of signal events while rejecting 97\% of background
events.
We see no significant difference between the selection efficiencies of
positive and negative candidates.

The reconstruction efficiency distribution over the Dalitz plot is modeled using
two-dimensional histograms formed from a sample of around $24\times10^{6}$
\BtoKPP\ MC events.
All selection criteria are applied except for those corresponding to the
invariant mass veto regions described below.
The ratio is taken of two histograms, the denominator containing the
true Dalitz-plot distribution of all generated MC events and the numerator
containing the reconstructed MC events.  The reconstructed events are
weighted in order to correct for differences between MC and data in the
particle identification and tracking efficiencies.
In order to give better resolution near the edges of the Dalitz plot, where most
reconstructed events lie, the histograms are formed in the ``square Dalitz
plot''~\cite{Aubert:2007jn} coordinates.
Linear interpolation is also applied between bins.
The efficiency shows very little variation across the majority of the
Dalitz plot
but decreases towards the corners where one of the particles has low
momentum.
The effect of experimental resolution on the signal model is neglected
since the resonances under consideration are sufficiently broad.
The average reconstruction efficiency for events in the signal box for the
nonresonant MC sample is 21.2\%.
As shown in~\eqref{SigDPLikeEqn}, in the likelihood fit we use
event-by-event efficiencies that depend on the Dalitz-plot position and are
different, but consistent, for \Bp\ and \Bm\ candidates.
The fraction of misreconstructed signal events is very small, $\sim 2\%$,
and so such events are not treated explicitly in the signal model.

\section{Backgrounds}

In addition to the continuum (\qqbar) background there are also backgrounds
from \BB\ events.  There are four main sources:
(i) combinatorial background from three unrelated tracks;
(ii) three- and four-body \B\ decays involving an intermediate $D$ meson;
(iii) charmless two- and four-body decays with an extra or missing
particle, and
(iv) three-body decays with one or more particles misidentified.
We veto candidates from charm and charmonium decays with large branching
fractions by rejecting events that have invariant masses (in units of
\gevcc) in the ranges:
$1.756 < m_{\Kp\pim} < 1.931$,
$1.660 < m_{\pipi}   < 1.800$,
$3.019 < m_{\pipi}   < 3.179$, and
$3.627 < m_{\pipi}   < 3.747$.
These ranges reject decays from \DzbtoKpi\ (or \pipi), \JPsitoll\ and
\Psitoll, respectively, where $\ell$ is a lepton that has been misidentified.

We study the remaining charm backgrounds that escape the vetoes and the
backgrounds from charmless \B\ decays with a large sample of MC-simulated
\BB\ decays equivalent to approximately 3 times the integrated
luminosity of the data sample.
Higher-statistics, exclusive MC samples are used to further study 68
\B-meson decay modes and to determine the \mes, \DeltaEpr, and Dalitz-plot
distributions that are used in the likelihood fit.
These distributions are normalized to the number of predicted events in the
final data sample, which we estimate using the reconstruction efficiencies
determined from the MC, the number of \BB\ pairs in our data sample, and the
branching fractions listed by the Particle Data Group~\cite{pdg2006} and the
Heavy Flavor Averaging Group~\cite{Barberio:2007cr}.
We further combine modes that have a similar behavior in each of the
discriminating variables \mes\ and \DeltaEpr\ into a \B-background category.
For each category combined Dalitz-plot, \mes, and \DeltaEpr\ PDFs are created,
and each is included as a separate component in the fit.
The predicted yields of \BB\ background events in the signal region are
$619\pm17$ $(659\pm18)$ for the negatively (positively) charged sample.

The background Dalitz-plot distributions are included in the likelihood fit through
the use of two-dimensional histograms.  For backgrounds from \B\ decays
these histograms are formed from the various MC samples.
For the continuum background the ``left sideband'' data sample is used.
This data sideband also contains events from \B\ decays and so MC samples 
are again used to subtract these events.
Like the reconstruction efficiency histograms, those for the backgrounds
are formed in the square Dalitz-plot coordinates and have linear interpolation applied
between bins.
Separate histograms are constructed for \Bp\ and \Bm\ events.
The \qqbar\ and \B-background PDFs are identical in their construction and
the \qqbar\ PDF is shown here as an example:
\begin{eqnarray}
\label{eq:BgDPLikeEqn}
&& \hspace{-10mm} {\cal L}_{\rm \qqbar}(\mACSq,\mBCSq) = \frac{1}{2} (1 - q_{\B} {\cal A}_{\qqbar}) \times \\ \nonumber
&& \left( \frac{\frac{1+q_{\B}}{2} \;
  Q(\mACSq,\mBCSq)}{\displaystyle{\int\!\!\int Q(\mACSq,\mBCSq) \; d\mACSq \,
    d\mBCSq} } + \right. \\ \nonumber
&& \phantom{\Bigg(++++} \left. \frac{\frac{1-q_{\B}}{2} \; \overline{Q}(\mACSq,\mBCSq)} {\displaystyle{\int\!\!\int \overline{Q}(\mACSq,\mBCSq) \; d\mACSq \, d\mBCSq}}\right)\,,
\end{eqnarray}
where ${\cal A}_{\qqbar}$ parametrizes possible charge asymmetry in the
background, and $Q(\mACSq,\mBCSq)$ and $\overline{Q}(\mACSq,\mBCSq)$ are
the Dalitz-plot distributions of the \qqbar\ events for \Bp\ and
\Bm\ events, respectively.

\section{Maximum Likelihood Fit}

To provide further discrimination between the signal and background
hypotheses in the likelihood fit we include PDFs for the kinematic
variables \mes\ and \DeltaEpr, which multiply that of the Dalitz plot.
The signal is modeled with a double Gaussian function in both cases.
The parameters of these functions are obtained from a sample of
\KPP\ MC events and are fixed in the fit to data.
The \qqbar\ \mes\ distribution is modeled with the experimentally motivated
ARGUS function~\cite{argus}. The end point for this ARGUS function is fixed to
$\sqrt{s}/2$ and the parameter describing the shape is fixed to the value
determined from the ``upper sideband.''
For \DeltaEpr\ the continuum is modeled with a linear function, the slope
of which is allowed to float in the fit to data.
The \BB\ background distributions are modeled using histograms obtained
from the mixture of \BB\ MC samples and are fixed in the fit.
The yields of signal and \qqbar\ events are allowed to float in the final fit
to the data while the yield of \BB\ background events is fixed.

The complete likelihood function is given by:
\begin{eqnarray}
{\cal L} &=& \exp\left(-\sum_k N_k\right) \times \\ \nonumber
&&
\prod_i^{N_e}
\Bigg[
\sum_k N_k {\cal P}_k^i(\mACSq,\mBCSq,\mes,\DeltaEpr,q_{\B})
\Bigg] \,,
\end{eqnarray}
where $N_k$ is the event yield for species $k$, $N_e$ is the total number
of events in the data sample, and ${\cal P}_k^i$ is the PDF for species $k$
for event $i$, which consists of a product of the Dalitz-plot, \mes, and \DeltaEpr\ PDFs.
The function $-\ln{\cal L}$ is used in the unbinned fit to the data.

We determine a nominal signal Dalitz-plot model using information from previous
studies~\cite{Abe:2002av,Aubert:2003mi,Garmash:2004wa,Aubert:2005ce,Garmash:2005rv} and the
change in the fit likelihood value observed when omitting or adding resonances.
In our previous study of \BtoKPP~\cite{Aubert:2005ce} 
we used a nominal model containing 
a phase-space nonresonant component and five intermediate resonant states:
\KstarIpip, \KpiSwavepip, \rhoIKp, \fIKp, and \chiczKp.
With the higher statistics and improved techniques of this analysis,
we find it necessary to include additional contributions from 
\KstarIIIpip, \fIIKp, and \fVIKp\ in order to achieve a reasonable
agreement of the fit with the data.

The first of these additions improves the agreement in 
the $\Kpm\pimp$ invariant mass projection;
although some discrepancy remains, this cannot be reduced by including other
known resonances nor by using alternative forms for the \KpiSwave\ shape.
In the $\pipm\pimp$ invariant mass projection,
we find the best agreement to data is achieved by including both the
\fIIKp\ and \fVIKp\ terms with ${\rm X} \!=\! 0$ ({\it i.e.} \fVI\ being a
scalar).
A reasonable fit can be achieved including 
instead a single broad vector resonance.
However, in the case ${\rm X} = 1$ it is natural to identify the 
\fVI\ as the \rhoII, and the observed large ratio of product branching fractions
between ${\cal B}(\BptorhoIIKp) \times {\cal B}(\rhoIItopippim)$ and 
${\cal B}(\BptorhoIKp) \times {\cal B}(\rhoItopippim)$
leads us to conclude that this cannot be the correct physical interpretation 
of the data.
Note that the \fIIKp\ contribution was observed by Belle~\cite{Garmash:2005rv}.

In addition, we include a small contribution from \omegaIKp, 
which is known to be present based on the well-measured
branching fractions of \BptoomegaIKp~\cite{Aubert:2007si,Jen:2006in} 
and \omegaItopippim~\cite{Akhmetshin:2006bx}.
Although the magnitude of this contribution is known to be very small,
due to the narrow width of the \omegaI, 
it can have a noticeable effect on the \rhoI\ line shape.
Thus, our nominal signal Dalitz-plot model comprises a phase-space nonresonant component
and nine intermediate resonance states:
\KstarIpip, \KpiSwavepip, \KstarIIIpip, \rhoIKp, \omegaIKp, \fIKp, \fIIKp,
\fVIKp, and \chiczKp.
This model differs from that used by Belle~\cite{Garmash:2005rv}
by the inclusion of \KstarIIIpip\ and the different parametrization of the 
\KpiSwavepip\ and nonresonant terms.

Our choice of parametrization of the \KpiSwavepip\ and nonresonant terms is
motivated by a number of physical considerations.  Different
parametrizations result in changes in likelihood that are highly dependent on
the composition of the rest of the decay model.  Compared to our final model,
including all resonant terms, the parametrization used by Belle gives a
larger fit likelihood.  However, this model exhibits large interference between
the \KstarII\ and nonresonant terms that can be either constructive or
destructive, leading to alternative solutions with similar likelihoods but
very different values of the fit fractions for these terms.  In the case that
these terms interfere destructively, the sum of fit fractions far exceeds
100\%, which we consider unlikely to be the correct physical description.
Moreover, the preferred solution can change depending on the exact composition
of the rest of the model -- in particular, we find that this problem, which
was previously reported by Belle~\cite{Garmash:2004wa}, is exacerbated by the
presence of the \KstarIII.  We have also tried a recent proposal for the
nonresonant distribution~\cite{Bediaga:2007mi} that results in a worse
likelihood. Therefore, we use in our nominal model the LASS parametrization
for the \KpiSwavepip\ and a nonresonant term as described above.

The mass and width of the \fVI\ are determined to be
$m_{f_{\rm X}} = (1479\pm8)\mevcc$, $\Gamma_{f_{\rm X}} = (80\pm19)\mev$,
with a correlation of $(-45\pm3)\%$, where the errors are statistical only
and are determined from a fit to the 2D likelihood profile.
These parameters are consistent with the values obtained by Belle~\cite{Garmash:2005rv}:
$m_{f_{\rm X}} = (1449\pm13)\mevcc$, $\Gamma_{f_{\rm X}} = (126\pm25)\mev$ 
(statistical errors only),
and with those listed for the $\fIV$~\cite{pdg2006}.

\begin{figure}[!htb]
\begin{center}
\includegraphics[angle=0, width=\columnwidth]{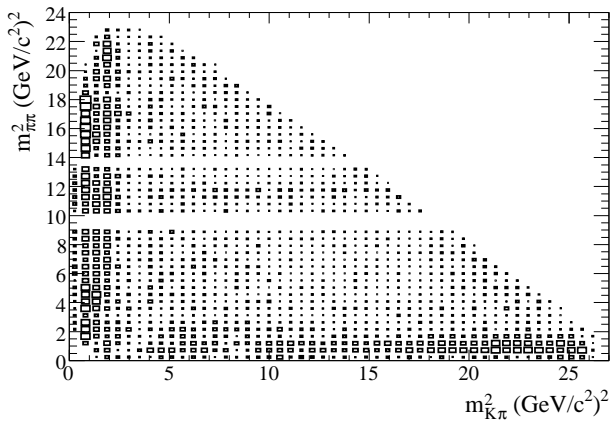}
\caption
{Background-subtracted Dalitz plot of the combined \BtoKPP\ data sample in the
signal region.  The plot shows bins with greater than zero entries.  The area
of the boxes is proportional to the number of entries.  The depleted
horizontal bands are the charmonium vetoes.}
\label{fig:dp}
\end{center}
\end{figure}

\begin{figure*}[!htb]
\begin{center}
\includegraphics[angle=0, width=0.49\textwidth]{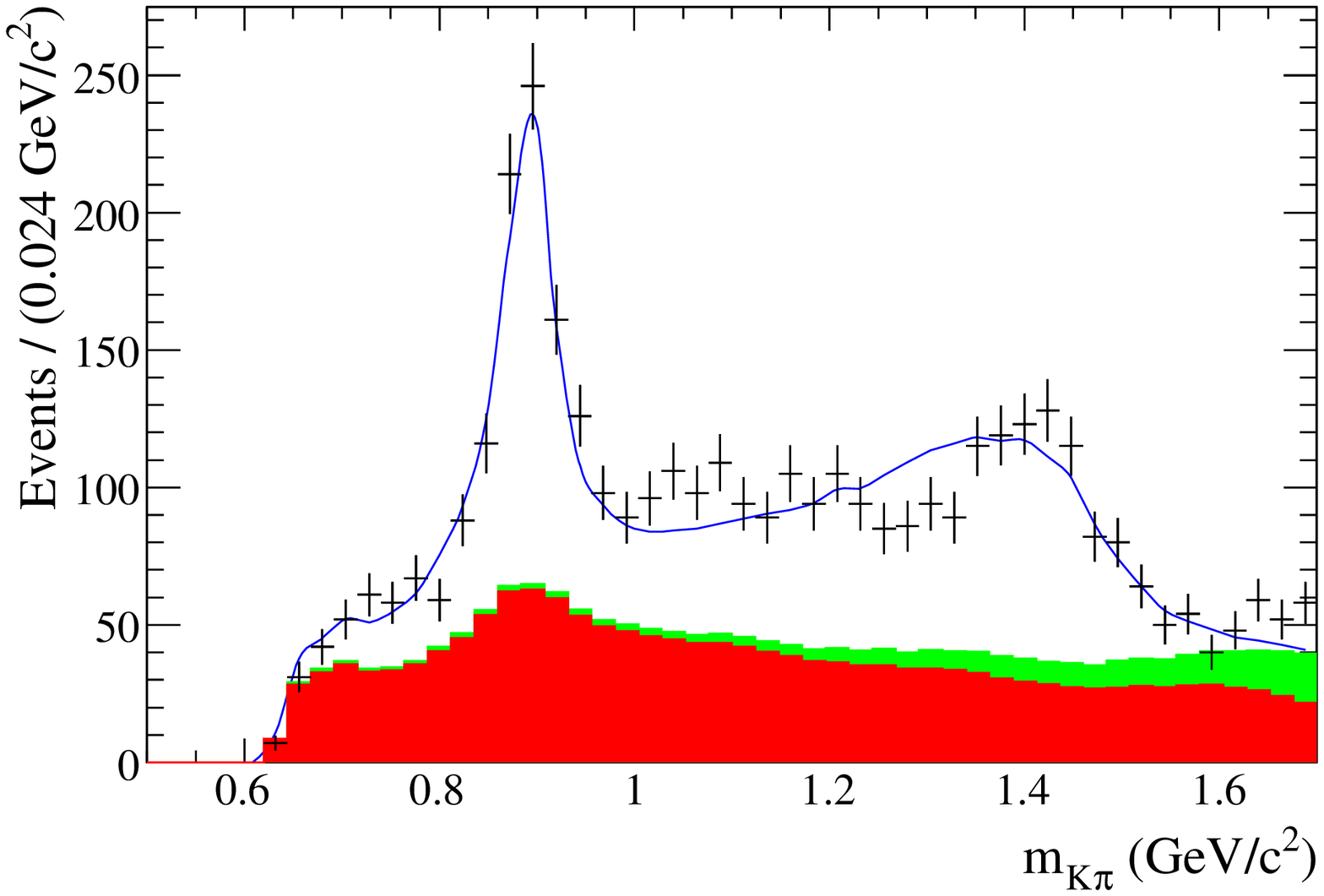}
\includegraphics[angle=0, width=0.49\textwidth]{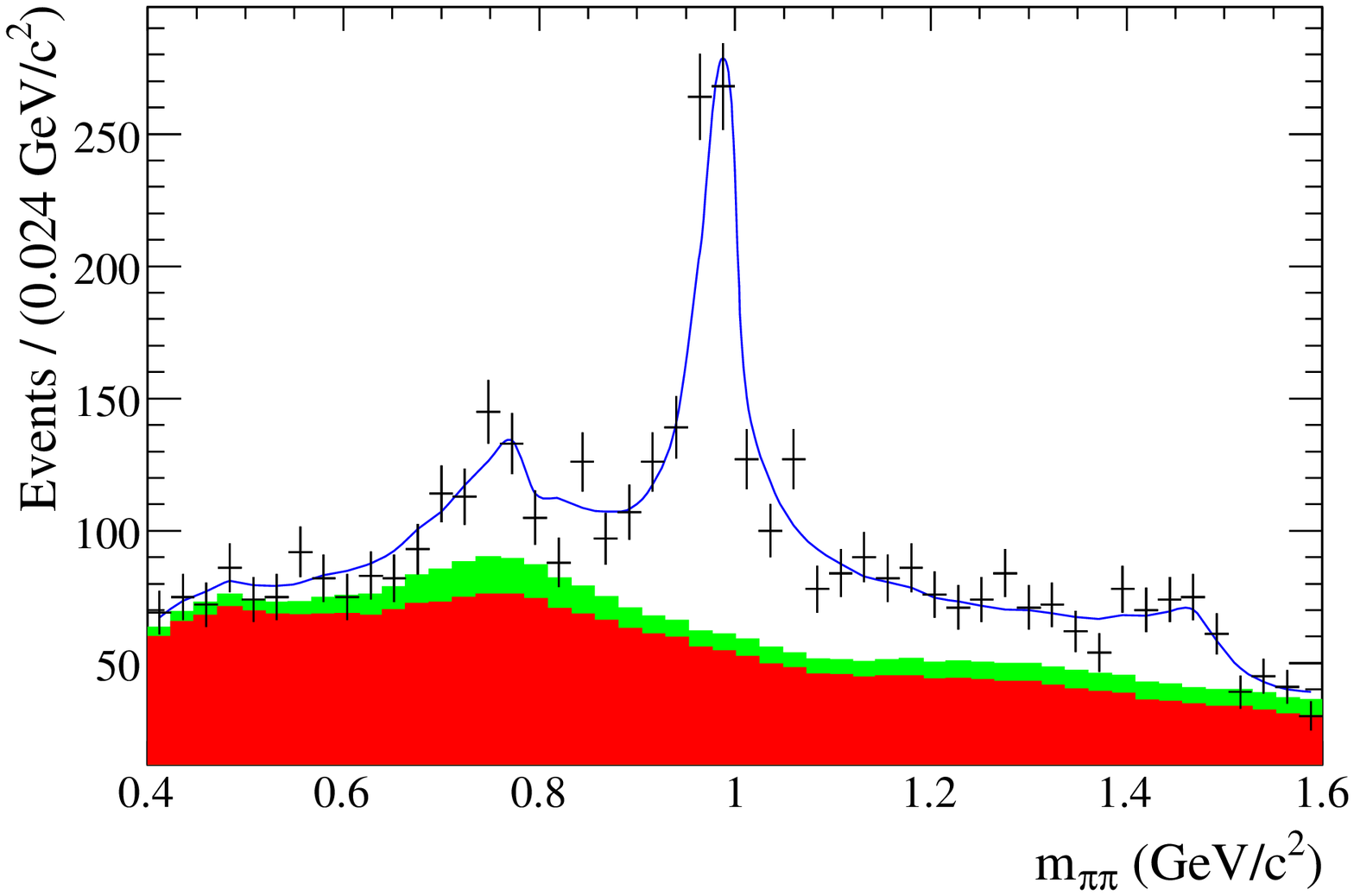}\\
\includegraphics[angle=0, width=0.49\textwidth]{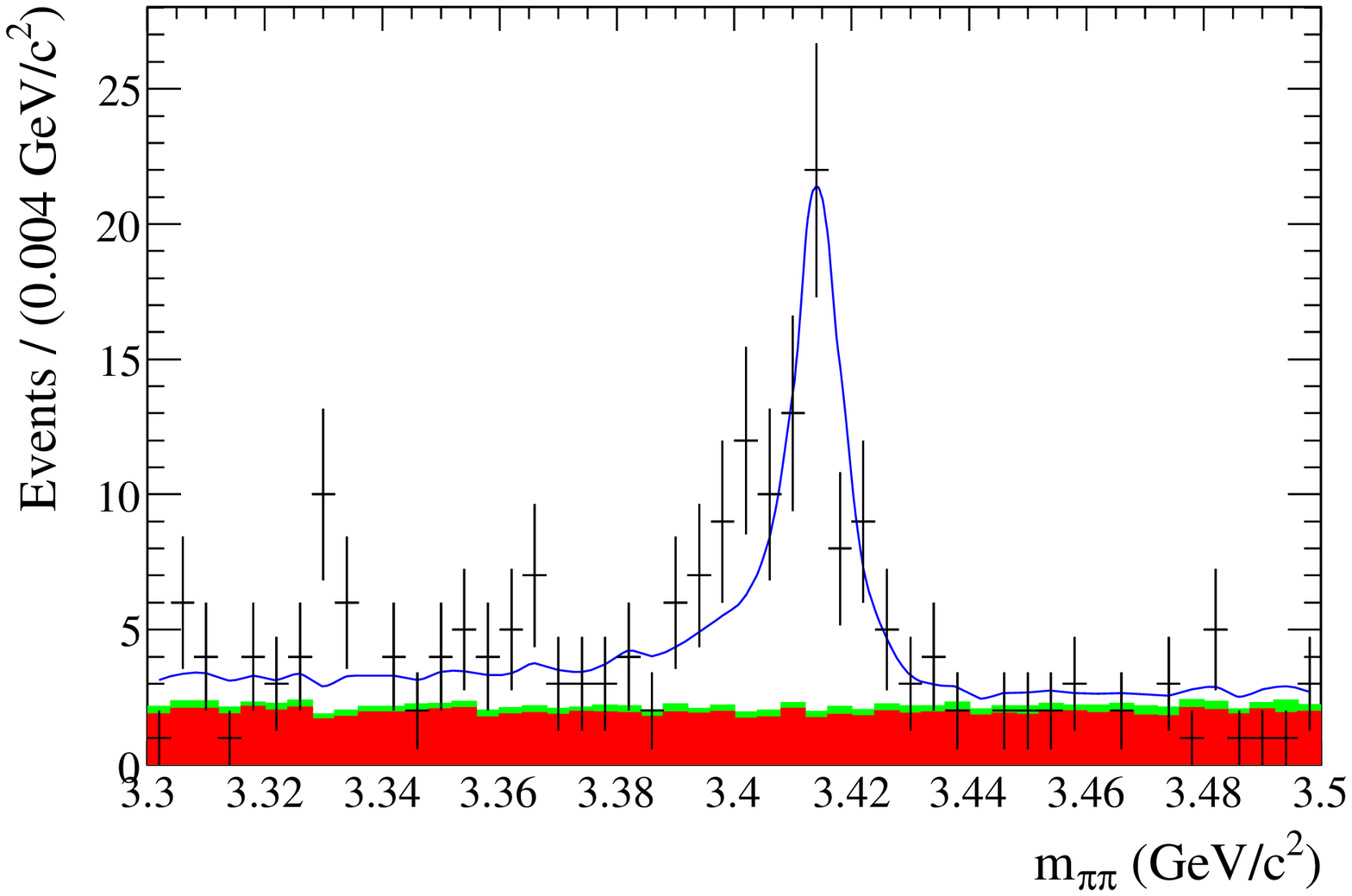}
\caption
{Invariant mass projections for the data in the signal region and the fit results.
 The left-hand plot shows the $m_{\Kpm\pimp}$ spectrum up to 1.7\gevcc.
 The right-hand plot shows the $m_{\pipm\pimp}$ spectrum up to 1.6\gevcc.
 The bottom plot shows the $m_{\pipm\pimp}$ spectrum in the region of
 the \chiczero.
 The data are the black points with statistical error bars,
 the lower solid (red/dark) histogram is the \qqbar\ component, 
 the middle solid (green/light) histogram is the \BB\ background contribution, 
 while the upper blue curve shows the total fit result. 
 For the $m_{\Kpm\pimp}$ plots the requirement is made that $m_{\pipm\pimp}$
 is greater than $2\gevcc$ and vice versa.}
\label{fig:nom}
\end{center}
\end{figure*}

We fit $12\,753$ signal candidates selected from the data using our nominal
Dalitz-plot model to obtain the central values of the $x_j$, $\Delta x_j$,
$y_j$, $\Delta y_j$ parameters for each component.
The $x$ and $y$ parameters of the \KstarI\ are fixed (to one and zero,
respectively) as the reference.
The \Dx\ and \Dy\ parameters of the \omegaI\ and nonresonant components are
fixed to zero in order to improve the fit stability.  They are allowed to
vary as a cross-check and are found to be consistent with zero.
The signal yield, \qqbar\ background yield, and asymmetry and the slope of
the \qqbar\ \DeltaEpr\ PDF are also floated parameters in the fit.
We then generate a large number of MC experiments using the fitted values,
and from the spread of results of fits to those experiments determine the
statistical uncertainties on those parameters, as well as the central
values and statistical uncertainties on the extracted parameters
${\it FF}_j$ and $\ACP_{\!,\,j}$.
This procedure takes into account correlations between the $x_j$, $\Delta
x_j$, $y_j$, $\Delta y_j$ parameters.
In order to make comparisons with previous measurements and predictions from
factorization models we multiply each fit fraction by the total branching
fraction to calculate the product branching fraction of each mode.
The results are shown in Tables~\ref{tab:summarytable}~and~\ref{tab:conclusion-results-summary}.
In order to determine the statistical significance of the direct
\CP\ violation exhibited by a component we evaluate the difference
$\Delta\ln{\cal L}$ between the negative log-likelihood of the nominal fit and
that of a fit where the $\Delta x$ and $\Delta y$ parameters for the given
component are fixed to zero.  This is then used to evaluate a $p$-value:
\begin{equation}
p = \int_{2\Delta\ln{\cal L}}^{\infty} f(z;n_d) \,dz \,,
\end{equation}
where $f(z;n_d)$ is the \chisq\ PDF and $n_d$ is the number of degrees of
freedom, two in this case.
We then determine the equivalent one-dimensional significance from this
$p$-value.
Note that this differs from the significance of $\ACP_{\!,\,j} \neq 0$,
since direct \CP\ violation can be observed in a Dalitz-plot analysis
not only through \B\ and \Bb\ amplitudes being different in magnitude,
but also by differences in their phases.
The significance estimations were cross-checked using MC simulations based
on the fit results.

The \KPP\ signal yield is found to be $4585\pm90\pm297\pm63$ and the total
charge asymmetry to be $(2.8\pm2.0\pm2.0\pm1.2)\%$, where the uncertainties
are statistical, systematic, and model-dependent, respectively.
The continuum background yield and charge asymmetry are found to be
$6830\pm110$ and $(-2.8\pm1.5)\%$, respectively, where the uncertainty is
statistical only.
The Dalitz plot of the data in the signal region, after subtraction of the
two background distributions (using the
\splot\ technique~\cite{Pivk:2004ty}), can be seen in \shortfigref{dp}.
Projections of the data, with the fit result overlaid, onto $\Kpm\pimp$
and $\pipm\pimp$ invariant mass distributions can be seen in
\shortfigref{nom}.
For the $m_{\Kpm\pimp}$ plots the requirement is made that $m_{\pipm\pimp}$ is
greater than $2\gevcc$ and vice versa in order to better illustrate the
structures present.
The agreement between the fit result and the data is generally very good,
although the discrepancy discussed above is visible in the $m_{\Kpm\pimp}$
spectrum.
Using the fitted signal distribution we calculate the average reconstruction
efficiency for our signal sample to be 22.5\%.

\renewcommand{\arraystretch}{1.5}

\begin{table*}
  \caption{Results of fits to data, with statistical, systematic, and model-dependent uncertainties.}
  \label{tab:summarytable}
\resizebox{\textwidth}{!}{
\begin{tabular}{l r@{$\pm$}c@{$\pm$}c@{$\,$}l r@{$\pm$}c@{$\pm$}c@{$\,$}l r@{$\pm$}c@{$\pm$}c@{$\,$}l r@{$\pm$}c@{$\pm$}c@{$\,$}l}
\hline
Resonance    & \fcc{$x$}                                           & \fcc{$y$}                                            & \fcc{\Dx}                                           & \fcc{\Dy}                                           \\
\hline
\KstarIpip   & \fcc{1.0 fixed}                                     & \fcc{0.0 fixed}                                      & $-0.017$ & $0.029$ & $0.005$ & $^{+0.004}_{-0.006}$ & $-0.238$ & $0.228$ & $0.062$ & $^{+0.144}_{-0.018}$ \\
\KpiSwavepip & $ 1.718$ & $0.084$ & $0.064$ & $^{+0.350}_{-0.055}$ & $-0.727$ & $ 0.108$ & $0.080$ & $^{+0.331}_{-0.111}$ & $-0.154$ & $0.131$ & $0.030$ & $^{+0.095}_{-0.010}$ & $-0.285$ & $0.337$ & $0.091$ & $^{+0.221}_{-0.019}$ \\
\rhoIKp      & $ 0.683$ & $0.075$ & $0.045$ & $^{+0.015}_{-0.073}$ & $-0.025$ & $ 0.135$ & $0.071$ & $^{+0.015}_{-0.073}$ & $-0.160$ & $0.049$ & $0.024$ & $^{+0.094}_{-0.013}$ & $ 0.169$ & $0.096$ & $0.057$ & $^{+0.133}_{-0.027}$ \\
\fIKp        & $-0.220$ & $0.200$ & $0.203$ & $^{+0.500}_{-0.095}$ & $ 1.203$ & $ 0.085$ & $0.052$ & $^{+0.113}_{-0.045}$ & $-0.109$ & $0.143$ & $0.087$ & $^{+0.037}_{-0.103}$ & $ 0.047$ & $0.045$ & $0.012$ & $^{+0.046}_{-0.018}$ \\
\chiczKp     & $-0.263$ & $0.044$ & $0.016$ & $^{+0.030}_{-0.014}$ & $ 0.180$ & $ 0.052$ & $0.034$ & $^{+0.225}_{-0.022}$ & $-0.033$ & $0.049$ & $0.012$ & $^{+0.017}_{-0.012}$ & $-0.007$ & $0.057$ & $0.019$ & $^{+0.006}_{-0.111}$ \\
Nonresonant  & $-0.594$ & $0.070$ & $0.170$ & $^{+0.112}_{-0.035}$ & $ 0.068$ & $ 0.132$ & $0.154$ & $^{+0.112}_{-0.099}$ & \fcc{0.0 fixed}                                     & \fcc{0.0 fixed}                                     \\
\KstarIIIpip & $-0.301$ & $0.060$ & $0.030$ & $^{+0.012}_{-0.134}$ & $ 0.424$ & $ 0.060$ & $0.045$ & $^{+0.012}_{-0.134}$ & $ 0.032$ & $0.078$ & $0.024$ & $^{+0.057}_{-0.050}$ & $ 0.007$ & $0.086$ & $0.017$ & $^{+0.025}_{-0.034}$ \\
\omegaIKp    & $-0.058$ & $0.067$ & $0.018$ & $^{+0.053}_{-0.011}$ & $ 0.100$ & $ 0.051$ & $0.010$ & $^{+0.033}_{-0.032}$ & \fcc{0.0 fixed}                                     & \fcc{0.0 fixed}                                     \\
\fIIKp       & $-0.193$ & $0.043$ & $0.022$ & $^{+0.026}_{-0.033}$ & $ 0.110$ & $ 0.050$ & $0.034$ & $^{+0.078}_{-0.073}$ & $-0.089$ & $0.046$ & $0.019$ & $^{+0.034}_{-0.014}$ & $ 0.125$ & $0.058$ & $0.021$ & $^{+0.034}_{-0.025}$ \\
\fVIKp       & $-0.290$ & $0.047$ & $0.064$ & $^{+0.047}_{-0.031}$ & $-0.136$ & $ 0.085$ & $0.098$ & $^{+0.102}_{-0.031}$ & $ 0.024$ & $0.040$ & $0.019$ & $^{+0.023}_{-0.018}$ & $ 0.056$ & $0.087$ & $0.044$ & $^{+0.010}_{-0.036}$ \\
\hline
\end{tabular}
}
\end{table*}

\renewcommand{\arraystretch}{1}

\section{Systematic Uncertainties}

The systematic uncertainties that affect the measurement of the fit
fractions, phases, and event yields are as follows.
The fixed \BB-background yields and asymmetries of the largest categories
are allowed to float and the variation of the other fitted parameters is
taken as the uncertainty.
The effect of the statistics of the MC and data sideband samples used to 
obtain the fixed shapes of the efficiency, \qqbar- and \BB-background
Dalitz-plot
histograms and the \BB-background histograms for \mes\ and \DeltaEpr\ is 
accounted for by fluctuating independently the histogram bin contents in
accordance with their errors and repeating the nominal fit.
The uncertainties on how well the samples model these distributions
are also taken into account through various cross-checks, including
variation of the charm veto range and comparison of continuum shapes
between sideband and signal region in MC samples.

The fixed parameters of the signal \mes\ and \DeltaEpr\ PDFs are studied in
the control sample \BptoDzbpip; \DzbtoKpi.  The parameters are determined
from MC and data samples, from which biases and scale factors are
calculated and used to adjust the parameters for the nominal fit.  The
parameters are then varied in accordance with the error on these biases and
scale factors and the fit repeated.
The uncertainties due to fixing the ARGUS parameter of the
\qqbar-background \mes\ PDF are determined by comparing the results of the
fit when the parameter value is obtained from off-peak data.

To confirm the fitting procedure, 500 MC experiments were performed in
which the events are generated from the PDFs used in the fit to data.
Small fit biases are observed for some of the fit parameters and are
included in the systematic uncertainties.
The contributions due to particle identification, tracking efficiency
corrections, and the calculation of \nbb\ are 4.2\%, 2.4\%, and 1.1\%,
respectively.
The efficiency correction due to the selection requirement on the NN has
also been calculated from \BptoDzbpip, \DzbtoKpi\ data and MC samples, and
is found to be $0.979\pm0.015$.  The error on this correction is
incorporated into the branching fraction systematic uncertainties.
Measured \CP\ asymmetries could be affected by detector charge bias.
In previous studies~\cite{Aubert:2005cp,Aubert:2007mj} this effect has
been estimated to be very small compared with the precision of our
measurements; we take it to be 0.5\%.

In addition to the above systematic uncertainties we also estimate
effects due to model-dependence, {\it i.e.} that characterize the
uncertainty on the results due to elements of the signal Dalitz-plot model.
The first of these elements consists of the parameters of the various
components of the signal model -- the masses and widths of all
resonances, the effective range and scattering length of the LASS model
of the \KpiSwave, the coupling constants of the \fI\ Flatt\'e
parametrization, and the value of the Blatt-Weisskopf barrier radius.
The associated uncertainties are evaluated by adjusting the parameters
within their experimental errors and refitting.
The second element is due to the different possible models both for the
nonresonant component, which is evaluated by refitting with the
parametrization used by Belle~\cite{Garmash:2005rv}, and for the \rhoI,
which is determined by refitting with the Gounaris-Sakurai form.
The third element is the uncertainty due to the composition of the signal
model.  It reflects observed changes in the parameters of the components
when the data are fitted with one of the smaller components removed from
the model and when the state \KstarIV\ is added to the model.
The uncertainties from each of these elements are added in quadrature to
obtain the final model-dependence.

\renewcommand{\arraystretch}{1.5}

\begin{table*}[htb]
\caption
{Summary of measurements of branching fractions (averaged over charge
conjugate states) and \CP\ asymmetries.
Note that these results are not corrected for secondary branching
fractions.
The first uncertainty is statistical, the second is systematic, and the
third represents the model dependence.
The final column is the statistical significance of direct \CP\ violation
determined as described in the text.
}
\label{tab:conclusion-results-summary}
\resizebox{\textwidth}{!}{
\begin{tabular}{lcccc}
\hline
Mode                           & Fit fraction (\%)                       & $\BR(\Bp \to {\rm Mode}) (10^{-6})$     & $A_{\CP}$ (\%)                      & DCPV sig.    \\
\hline
\Kpppos\ total                 &                                         & $54.4\pm1.1\pm4.5\pm0.7$                & $2.8\pm2.0\pm2.0\pm1.2$             &              \\
\hline                                                                                                                                                            
\KstarIpip; \KstarItoKppim     & $13.3\pm0.7\pm0.7\,^{+0.4}_{-0.9}$      & $7.2\pm0.4\pm0.7\,^{+0.3}_{-0.5}$       & $+3.2\pm5.2\pm1.1\,^{+1.2}_{-0.7}$  &  $0.9\sigma$ \\
\KpiSwavepip; \KpiSwavetoKppim & $45.0\pm1.4\pm1.2\,^{+12.9}_{-0.2}$     & $24.5\pm0.9\pm2.1\,^{+7.0}_{-1.1}$      & $+3.2\pm3.5\pm2.0\,^{+2.7}_{-1.9}$  &  $1.2\sigma$ \\
\rhoIKp; \rhoItopippim         & $6.54\pm0.81\pm0.58\,^{+0.69}_{-0.26}$  & $3.56\pm0.45\pm0.43\,^{+0.38}_{-0.15}$  & $+44\pm10\pm4\,^{+5}_{-13}$         &  $3.7\sigma$ \\
\fIKp; \fItopippim             & $18.9\pm0.9\pm1.7\,^{+2.8}_{-0.6}$      & $10.3\pm0.5\pm1.3\,^{+1.5}_{-0.4}$      & $-10.6\pm5.0\pm1.1\,^{+3.4}_{-1.0}$ &  $1.8\sigma$ \\
\chiczKp; \chicztopippim       & $1.29\pm0.19\pm0.15\,^{+0.12}_{-0.03}$  & $0.70\pm0.10\pm0.10\,^{+0.06}_{-0.02}$  & $-14\pm15\pm3\,^{+1}_{-5}$          &  $0.5\sigma$ \\
\NonRes                        & $4.5\pm0.9\pm2.4\,^{+0.6}_{-1.5}$       & $2.4\pm0.5\pm1.3\,^{+0.3}_{-0.8}$       & ---                                 &  ---         \\
\KstarIIIpip; \KstarIIItoKppim & $3.40\pm0.75\pm0.42\,^{+0.99}_{-0.13}$  & $1.85\pm0.41\pm0.28\,^{+0.54}_{-0.08}$  & $+5\pm23\pm4\,^{+18}_{-7}$          &  $0.2\sigma$ \\
\omegaIKp; \omegaItopippim     & $0.17\pm0.24\pm0.03\,^{+0.05}_{-0.08}$  & $0.09\pm0.13\pm0.02\,^{+0.03}_{-0.04}$  & ---                                 &  ---         \\
\fIIKp; \fIItopippim           & $0.91\pm0.27\pm0.11\,^{+0.24}_{-0.17}$  & $0.50\pm0.15\pm0.07\,^{+0.13}_{-0.09}$  & $-85\pm22\pm13\,^{+22}_{-2}$        &  $3.5\sigma$ \\
\fVIKp; \fVItopippim           & $1.33\pm0.38\pm0.86\,^{+0.04}_{-0.14}$  & $0.73\pm0.21\pm0.47\,^{+0.02}_{-0.08}$  & $+28\pm26\pm13\,^{+7}_{-5}$         &  $0.6\sigma$ \\
\hline
\end{tabular}
}
\end{table*}

\renewcommand{\arraystretch}{1}

\section{Summary and Discussion}

Our results are shown in Tables~\ref{tab:summarytable}~and~\ref{tab:conclusion-results-summary}.
The total branching fraction
${\cal B}(\BtoKpppos) = (54.4\pm1.1\pm4.5\pm0.7)\times10^{-6}$ 
is compatible with Belle's measurement of
$(48.8\pm1.1\pm3.6)\times10^{-6}$~\cite{Garmash:2005rv}.
This result was cross-checked by using the same procedure to measure the
\BptoDzbpip; \DzbtoKpi\ branching fraction, which was found to be consistent
with the PDG value~\cite{pdg2006}.
The total charge asymmetry for \BtoKpppos\ has been measured to be
consistent with zero to a higher degree of precision than previous
measurements.

We see evidence of large direct \CP\ violation in \BptorhoIKp, consistent
with the findings of our previous analysis~\cite{Aubert:2005ce} and of
Belle~\cite{Garmash:2005rv}.
The statistical significance of the direct \CP\ violation effect is found to
be $3.7\sigma$ from the change in likelihood when the \Dx\ and \Dy\ terms
associated with $\rhoIKp$ are fixed to zero.  We have verified this estimate
of the significance using MC simulations.  As experimental systematic
uncertainties are much smaller than the statistical errors, they do not affect
this conclusion.  We have cross-checked the effect of the choice of the Dalitz
model on the significance.  We find that the significance remains above
$3\sigma$ with alternative models, including that used by Belle~\cite{Garmash:2005rv}.

\begin{figure}[!htb]
\begin{center}
\includegraphics[angle=0, width=\columnwidth]{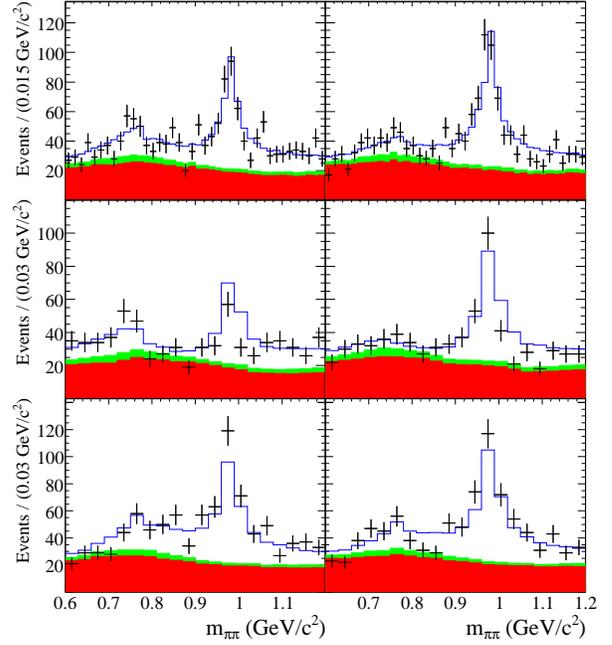}
\caption
{
Projection plots of the \pipi\ invariant mass in the region of the
\rhoI\ and \fI\ resonances.
The left (right) plots are for \Bm\ (\Bp) candidates.
The top row shows all candidates, the middle row shows those where
$\cos\theta_H>0$, and the bottom row shows those where $\cos\theta_H<0$.
The data are the black points with statistical error bars,
the lower solid (red/dark) histogram is the \qqbar\ component, 
the middle solid (green/light) histogram is the \BB\ background contribution, 
while the blue open histogram shows the total fit result. 
}
\label{fig:rho-projs}
\end{center}
\end{figure}

Plots of the \pipi\ invariant mass projections in the region of the
\rhoI\ and \fI\ are shown separately for \Bp\ and \Bm\ candidates in
\shortfigref{rho-projs}.
In an attempt to highlight the direct \CP\ violation effect, we also show
plots with the data further subdivided into positive and negative values of
$\cos\theta_H = \vec{p}\cdot\vec{q}\,/\!\left(|\vec{p}\,||\vec{q}\,|\right)$,
using the notation of \eqref{angular}.
The asymmetry in the excess of events above background in the \rhoI\ region
is particularly apparent in the distributions with the requirement
$\cos\theta_H>0$.

The statistical significance of direct \CP\ violation in \BptofIIKp\
is also above $3\sigma$, but this result suffers from large model uncertainties.
The \KstarIpip, \KpiSwavepip, and \KstarIIIpip\ charge asymmetries are all
consistent with zero, as expected.  

After correcting for the secondary branching fraction
$\BR(\KstarItoKppim)=\frac{2}{3}$ we find $\BR(\BptoKstarIpip)$
to be
$(10.8\pm0.6\pm1.1\,^{+0.4}_{-0.8})\times10^{-6}$.
Similarly we find
$\BR(\BptofIIKp)$ to be
$(0.88\pm0.26\pm0.13\,^{+0.23}_{-0.16}\,^{+0.01}_{-0.02})\times10^{-6}$ and
$\BR(\BptoKstarIIIpip)$ to be
$(5.6\pm1.2\pm0.8\,^{+1.6}_{-0.2}\pm0.1)\times10^{-6}$,
where the fourth errors are due to the uncertainties on the secondary
branching fractions.
These and the other branching fraction measurements are, in general,
consistent with previous measurements.
The product branching fraction of \BptoomegaIKp; \omegaItopippim\ agrees
with the expectation from previous measurements, albeit with large
uncertainties.
The only disparities between our results and those reported by
Belle~\cite{Garmash:2005rv} arise from the different treatments of the
\Kpi\ S-wave.
The forward-backward asymmetry apparent in both the \KstarI\ and \fI\ bands in
\shortfigref{dp} is well reproduced by the fit and is due to S--P-wave
interference in the Dalitz plot.

The \KpiSwave\ component is modeled in our analysis by the LASS
parametrization~\cite{LASS}, which consists of a nonresonant effective
range term plus a relativistic Breit-Wigner term for the
\KstarII\ resonance itself.
This parametrization makes use of the available experimental information,
however the size of the phase space here means that the range of
\Kpi\ invariant masses under consideration is much greater than in previous
studies.  The agreement between the model and the data in the region of the
\KstarII\ is not as good as in the remainder of the Dalitz plot but all
alternative models tried yielded poorer results.
If we assume that the model used is correct then we can calculate the
branching fraction for \BptoKstarIIpip\ and find it to be:
$(32.0\pm1.2\pm2.7\,^{+9.1}_{-1.4}\pm5.2)\times10^{-6}$,
where the fourth error is due to the uncertainty on the branching fraction of
$\KstarII \to K\pi$ combined with the uncertainty on the proportion of the
\KpiSwave\ component due to the \KstarII\ resonance.
In addition we can calculate the total nonresonant contribution by
combining coherently the nonresonant part of the LASS parametrization and
the phase-space nonresonant.  We find this total nonresonant branching
fraction to be:
$(9.3\pm1.0\pm1.2\,^{+6.7}_{-0.4}\pm1.2)\times10^{-6}$,
where the fourth error is due to the uncertainty on the proportion of the
\KpiSwave\ component that is nonresonant.
The Belle collaboration finds somewhat larger \KstarII\ and nonresonant
branching fractions though they treat the \KstarII\ as a Breit-Wigner
component, separate from the rest of the S-wave modeled as a nonresonant
amplitude that has variation in magnitude but no variation in phase over
the Dalitz plot~\cite{Garmash:2005rv}.

In conclusion, we have performed a Dalitz-plot analysis of 
\BtoKPP\ decays based on a \onreslumi\ data sample containing
\bbpairs\ \BB\ pairs collected by the \babar\ detector.
To obtain a good fit to the data we find that contributions from
\fIIKp\ and \fVIKp\ are necessary, with \fVI\ being a scalar with parameters
$m_{f_{\rm X}} = (1479\pm8)\mevcc$, $\Gamma_{f_{\rm X}} = (80\pm19)\mev$, with a
correlation of $(-45\pm3)\%$, where the errors are statistical only.
These are consistent with the mass and width of the \fIV.
In the $\Kpm\pimp$ invariant mass projection some discrepancy remains 
in the $1200$--$1400\mevcc$ range --
our model includes \KpiSwavepip\ (using the LASS parametrization) and 
\KstarIIIpip\ terms in this region, although some alternative models also give
a good description of the data.
We measure \CP-averaged branching fractions and direct \CP\ asymmetries
for intermediate resonant and nonresonant contributions.
Our results are consistent with the standard model, and can be used
together with results from other $B \to K\pi\pi$ decays to obtain
constraints on the CKM phase $\gamma$.
We find evidence for direct \CP\ violation in the decay \BtorhoIK,
with a \CP-violation parameter $A_{\CP} = (+44\pm10\pm4\,^{+5}_{-13})\%$ .
These results supersede those in our previous publication~\cite{Aubert:2005ce}.

\section{Acknowledgements}

We are grateful for the 
extraordinary contributions of our \pep2\ colleagues in
achieving the excellent luminosity and machine conditions
that have made this work possible.
The success of this project also relies critically on the 
expertise and dedication of the computing organizations that 
support \babar.
The collaborating institutions wish to thank 
SLAC for its support and the kind hospitality extended to them. 
This work is supported by the
US Department of Energy
and National Science Foundation, the
Natural Sciences and Engineering Research Council (Canada),
the Commissariat \`a l'Energie Atomique and
Institut National de Physique Nucl\'eaire et de Physique des Particules
(France), the
Bundesministerium f\"ur Bildung und Forschung and
Deutsche Forschungsgemeinschaft
(Germany), the
Istituto Nazionale di Fisica Nucleare (Italy),
the Foundation for Fundamental Research on Matter (The Netherlands),
the Research Council of Norway, the
Ministry of Education and Science of the Russian Federation, 
Ministerio de Educaci\'on y Ciencia (Spain), and the
Science and Technology Facilities Council (United Kingdom).
Individuals have received support from 
the Marie-Curie IEF program (European Union) and
the A. P. Sloan Foundation.

\bibliography{references}

\begin{thebibliography}{75}
\expandafter\ifx\csname natexlab\endcsname\relax\def\natexlab#1{#1}\fi
\expandafter\ifx\csname bibnamefont\endcsname\relax
  \def\bibnamefont#1{#1}\fi
\expandafter\ifx\csname bibfnamefont\endcsname\relax
  \def\bibfnamefont#1{#1}\fi
\expandafter\ifx\csname citenamefont\endcsname\relax
  \def\citenamefont#1{#1}\fi
\expandafter\ifx\csname url\endcsname\relax
  \def\url#1{\texttt{#1}}\fi
\expandafter\ifx\csname urlprefix\endcsname\relax\def\urlprefix{URL }\fi
\providecommand{\bibinfo}[2]{#2}
\providecommand{\eprint}[2][]{\url{#2}}

\bibitem[{\citenamefont{Kobayashi and Maskawa}(1973)}]{Kobayashi:1973fv}
\bibinfo{author}{\bibfnamefont{M.}~\bibnamefont{Kobayashi}} \bibnamefont{and}
  \bibinfo{author}{\bibfnamefont{T.}~\bibnamefont{Maskawa}},
  \bibinfo{journal}{Prog. Theor. Phys.} \textbf{\bibinfo{volume}{49}},
  \bibinfo{pages}{652} (\bibinfo{year}{1973}).

\bibitem[{\citenamefont{Alavi-Harati et~al.}(1999)}]{AlaviHarati:1999xp}
\bibinfo{author}{\bibfnamefont{A.}~\bibnamefont{Alavi-Harati}}
  \bibnamefont{et~al.} (\bibinfo{collaboration}{KTeV}), \bibinfo{journal}{Phys.
  Rev. Lett.} \textbf{\bibinfo{volume}{83}}, \bibinfo{pages}{22}
  (\bibinfo{year}{1999}).

\bibitem[{\citenamefont{Fanti et~al.}(1999)}]{Fanti:1999nm}
\bibinfo{author}{\bibfnamefont{V.}~\bibnamefont{Fanti}} \bibnamefont{et~al.}
  (\bibinfo{collaboration}{NA48}), \bibinfo{journal}{Phys. Lett.}
  \textbf{\bibinfo{volume}{B465}}, \bibinfo{pages}{335} (\bibinfo{year}{1999}).

\bibitem[{\citenamefont{Aubert et~al.}(2001)}]{Aubert:2001nu}
\bibinfo{author}{\bibfnamefont{B.}~\bibnamefont{Aubert}} \bibnamefont{et~al.}
  (\bibinfo{collaboration}{\babar}), \bibinfo{journal}{Phys. Rev. Lett.}
  \textbf{\bibinfo{volume}{87}}, \bibinfo{pages}{091801}
  (\bibinfo{year}{2001}).

\bibitem[{\citenamefont{Abe et~al.}(2001)}]{Abe:2001xe}
\bibinfo{author}{\bibfnamefont{K.}~\bibnamefont{Abe}} \bibnamefont{et~al.}
  (\bibinfo{collaboration}{Belle}), \bibinfo{journal}{Phys. Rev. Lett.}
  \textbf{\bibinfo{volume}{87}}, \bibinfo{pages}{091802}
  (\bibinfo{year}{2001}).

\bibitem[{\citenamefont{Aubert et~al.}(2004{\natexlab{a}})}]{Aubert:2004qm}
\bibinfo{author}{\bibfnamefont{B.}~\bibnamefont{Aubert}} \bibnamefont{et~al.}
  (\bibinfo{collaboration}{\babar}), \bibinfo{journal}{Phys. Rev. Lett.}
  \textbf{\bibinfo{volume}{93}}, \bibinfo{pages}{131801}
  (\bibinfo{year}{2004}{\natexlab{a}}).

\bibitem[{\citenamefont{Yao et~al.}(2006)}]{pdg2006}
\bibinfo{author}{\bibfnamefont{W.~M.} \bibnamefont{Yao}} \bibnamefont{et~al.}
  (\bibinfo{collaboration}{Particle Data Group}), \bibinfo{journal}{J. Phys.}
  \textbf{\bibinfo{volume}{G33}}, \bibinfo{pages}{1} (\bibinfo{year}{2006}).

\bibitem[{\citenamefont{Barberio et~al.}(2007)}]{Barberio:2007cr}
\bibinfo{author}{\bibfnamefont{E.}~\bibnamefont{Barberio}} \bibnamefont{et~al.}
  (\bibinfo{collaboration}{Heavy Flavor Averaging Group (HFAG)})
  (\bibinfo{year}{2007}), \eprint{arXiv:0704.3575 [hep-ex]}.

\bibitem[{\citenamefont{Aubert et~al.}(2007{\natexlab{a}})}]{Aubert:2007hh}
\bibinfo{author}{\bibfnamefont{B.}~\bibnamefont{Aubert}} \bibnamefont{et~al.}
  (\bibinfo{collaboration}{\babar}), \bibinfo{journal}{Phys. Rev.}
  \textbf{\bibinfo{volume}{D76}}, \bibinfo{pages}{091102}
  (\bibinfo{year}{2007}{\natexlab{a}}).

\bibitem[{\citenamefont{Lin et~al.}(2008)}]{Lin:2008zz}
\bibinfo{author}{\bibfnamefont{S.~W.} \bibnamefont{Lin}} \bibnamefont{et~al.}
  (\bibinfo{collaboration}{Belle}), \bibinfo{journal}{Nature}
  \textbf{\bibinfo{volume}{452}}, \bibinfo{pages}{332} (\bibinfo{year}{2008}).

\bibitem[{\citenamefont{Buras et~al.}(2004)\citenamefont{Buras, Fleischer,
  Recksiegel, and Schwab}}]{Buras:2003dj}
\bibinfo{author}{\bibfnamefont{A.~J.} \bibnamefont{Buras}},
  \bibinfo{author}{\bibfnamefont{R.}~\bibnamefont{Fleischer}},
  \bibinfo{author}{\bibfnamefont{S.}~\bibnamefont{Recksiegel}},
  \bibnamefont{and} \bibinfo{author}{\bibfnamefont{F.}~\bibnamefont{Schwab}},
  \bibinfo{journal}{Phys. Rev. Lett.} \textbf{\bibinfo{volume}{92}},
  \bibinfo{pages}{101804} (\bibinfo{year}{2004}).

\bibitem[{\citenamefont{Buras et~al.}(2006)\citenamefont{Buras, Fleischer,
  Recksiegel, and Schwab}}]{Buras:2005cv}
\bibinfo{author}{\bibfnamefont{A.~J.} \bibnamefont{Buras}},
  \bibinfo{author}{\bibfnamefont{R.}~\bibnamefont{Fleischer}},
  \bibinfo{author}{\bibfnamefont{S.}~\bibnamefont{Recksiegel}},
  \bibnamefont{and} \bibinfo{author}{\bibfnamefont{F.}~\bibnamefont{Schwab}},
  \bibinfo{journal}{Eur. Phys. J.} \textbf{\bibinfo{volume}{C45}},
  \bibinfo{pages}{701} (\bibinfo{year}{2006}).

\bibitem[{\citenamefont{Baek et~al.}(2005)\citenamefont{Baek, Hamel, London,
  Datta, and Suprun}}]{Baek:2004rp}
\bibinfo{author}{\bibfnamefont{S.}~\bibnamefont{Baek}},
  \bibinfo{author}{\bibfnamefont{P.}~\bibnamefont{Hamel}},
  \bibinfo{author}{\bibfnamefont{D.}~\bibnamefont{London}},
  \bibinfo{author}{\bibfnamefont{A.}~\bibnamefont{Datta}}, \bibnamefont{and}
  \bibinfo{author}{\bibfnamefont{D.~A.} \bibnamefont{Suprun}},
  \bibinfo{journal}{Phys. Rev.} \textbf{\bibinfo{volume}{D71}},
  \bibinfo{pages}{057502} (\bibinfo{year}{2005}).

\bibitem[{\citenamefont{Kim et~al.}(2005)\citenamefont{Kim, Oh, and
  Yu}}]{Kim:2005jp}
\bibinfo{author}{\bibfnamefont{C.~S.} \bibnamefont{Kim}},
  \bibinfo{author}{\bibfnamefont{S.}~\bibnamefont{Oh}}, \bibnamefont{and}
  \bibinfo{author}{\bibfnamefont{C.}~\bibnamefont{Yu}}, \bibinfo{journal}{Phys.
  Rev.} \textbf{\bibinfo{volume}{D72}}, \bibinfo{pages}{074005}
  (\bibinfo{year}{2005}).

\bibitem[{\citenamefont{Li and Yang}(2005)}]{Li:2005wx}
\bibinfo{author}{\bibfnamefont{X.-Q.} \bibnamefont{Li}} \bibnamefont{and}
  \bibinfo{author}{\bibfnamefont{Y.-D.} \bibnamefont{Yang}},
  \bibinfo{journal}{Phys. Rev.} \textbf{\bibinfo{volume}{D72}},
  \bibinfo{pages}{074007} (\bibinfo{year}{2005}).

\bibitem[{\citenamefont{Li et~al.}(2005)\citenamefont{Li, Mishima, and
  Sanda}}]{Li:2005kt}
\bibinfo{author}{\bibfnamefont{H.-N.} \bibnamefont{Li}},
  \bibinfo{author}{\bibfnamefont{S.}~\bibnamefont{Mishima}}, \bibnamefont{and}
  \bibinfo{author}{\bibfnamefont{A.~I.} \bibnamefont{Sanda}},
  \bibinfo{journal}{Phys. Rev.} \textbf{\bibinfo{volume}{D72}},
  \bibinfo{pages}{114005} (\bibinfo{year}{2005}).

\bibitem[{\citenamefont{Chen}(2002)}]{Chen:2001jx}
\bibinfo{author}{\bibfnamefont{C.-H.} \bibnamefont{Chen}},
  \bibinfo{journal}{Phys. Lett.} \textbf{\bibinfo{volume}{B525}},
  \bibinfo{pages}{56} (\bibinfo{year}{2002}).

\bibitem[{\citenamefont{Du et~al.}(2002)\citenamefont{Du, Gong, Sun, Yang, and
  Zhu}}]{Du:2002up}
\bibinfo{author}{\bibfnamefont{D.-S.} \bibnamefont{Du}},
  \bibinfo{author}{\bibfnamefont{H.-J.} \bibnamefont{Gong}},
  \bibinfo{author}{\bibfnamefont{J.-F.} \bibnamefont{Sun}},
  \bibinfo{author}{\bibfnamefont{D.-S.} \bibnamefont{Yang}}, \bibnamefont{and}
  \bibinfo{author}{\bibfnamefont{G.-H.} \bibnamefont{Zhu}},
  \bibinfo{journal}{Phys. Rev.} \textbf{\bibinfo{volume}{D65}},
  \bibinfo{pages}{094025} (\bibinfo{year}{2002}).

\bibitem[{\citenamefont{Beneke and Neubert}(2003)}]{Beneke:2003zv}
\bibinfo{author}{\bibfnamefont{M.}~\bibnamefont{Beneke}} \bibnamefont{and}
  \bibinfo{author}{\bibfnamefont{M.}~\bibnamefont{Neubert}},
  \bibinfo{journal}{Nucl. Phys.} \textbf{\bibinfo{volume}{B675}},
  \bibinfo{pages}{333} (\bibinfo{year}{2003}).

\bibitem[{\citenamefont{Chiang et~al.}(2004)\citenamefont{Chiang, Gronau, Luo,
  Rosner, and Suprun}}]{Chiang:2003pm}
\bibinfo{author}{\bibfnamefont{C.-W.} \bibnamefont{Chiang}},
  \bibinfo{author}{\bibfnamefont{M.}~\bibnamefont{Gronau}},
  \bibinfo{author}{\bibfnamefont{Z.}~\bibnamefont{Luo}},
  \bibinfo{author}{\bibfnamefont{J.~L.} \bibnamefont{Rosner}},
  \bibnamefont{and} \bibinfo{author}{\bibfnamefont{D.~A.}
  \bibnamefont{Suprun}}, \bibinfo{journal}{Phys. Rev.}
  \textbf{\bibinfo{volume}{D69}}, \bibinfo{pages}{034001}
  (\bibinfo{year}{2004}).

\bibitem[{\citenamefont{Isola et~al.}(2003)\citenamefont{Isola, Ladisa,
  Nardulli, and Santorelli}}]{Isola:2003fh}
\bibinfo{author}{\bibfnamefont{C.}~\bibnamefont{Isola}},
  \bibinfo{author}{\bibfnamefont{M.}~\bibnamefont{Ladisa}},
  \bibinfo{author}{\bibfnamefont{G.}~\bibnamefont{Nardulli}}, \bibnamefont{and}
  \bibinfo{author}{\bibfnamefont{P.}~\bibnamefont{Santorelli}},
  \bibinfo{journal}{Phys. Rev.} \textbf{\bibinfo{volume}{D68}},
  \bibinfo{pages}{114001} (\bibinfo{year}{2003}).

\bibitem[{\citenamefont{Li and Yang}(2006)}]{Li:2006jb}
\bibinfo{author}{\bibfnamefont{X.-Q.} \bibnamefont{Li}} \bibnamefont{and}
  \bibinfo{author}{\bibfnamefont{Y.-D.} \bibnamefont{Yang}},
  \bibinfo{journal}{Phys. Rev.} \textbf{\bibinfo{volume}{D73}},
  \bibinfo{pages}{114027} (\bibinfo{year}{2006}).

\bibitem[{\citenamefont{Li and Mishima}(2006)}]{Li:2006jv}
\bibinfo{author}{\bibfnamefont{H.-N.} \bibnamefont{Li}} \bibnamefont{and}
  \bibinfo{author}{\bibfnamefont{S.}~\bibnamefont{Mishima}},
  \bibinfo{journal}{Phys. Rev.} \textbf{\bibinfo{volume}{D74}},
  \bibinfo{pages}{094020} (\bibinfo{year}{2006}).

\bibitem[{\citenamefont{El-Bennich et~al.}(2006)\citenamefont{El-Bennich,
  Furman, Kaminski, Lesniak, and Loiseau}}]{ElBennich:2006yi}
\bibinfo{author}{\bibfnamefont{B.}~\bibnamefont{El-Bennich}},
  \bibinfo{author}{\bibfnamefont{A.}~\bibnamefont{Furman}},
  \bibinfo{author}{\bibfnamefont{R.}~\bibnamefont{Kaminski}},
  \bibinfo{author}{\bibfnamefont{L.}~\bibnamefont{Lesniak}}, \bibnamefont{and}
  \bibinfo{author}{\bibfnamefont{B.}~\bibnamefont{Loiseau}},
  \bibinfo{journal}{Phys. Rev.} \textbf{\bibinfo{volume}{D74}},
  \bibinfo{pages}{114009} (\bibinfo{year}{2006}).

\bibitem[{\citenamefont{Wang et~al.}(2008)\citenamefont{Wang, Wang, Yang, and
  Lu}}]{Wang:2008rk}
\bibinfo{author}{\bibfnamefont{W.}~\bibnamefont{Wang}},
  \bibinfo{author}{\bibfnamefont{Y.-M.} \bibnamefont{Wang}},
  \bibinfo{author}{\bibfnamefont{D.-S.} \bibnamefont{Yang}}, \bibnamefont{and}
  \bibinfo{author}{\bibfnamefont{C.-D.} \bibnamefont{Lu}}
  (\bibinfo{year}{2008}), \eprint{arXiv:0801.3123 [hep-ph]}.

\bibitem[{\citenamefont{Aubert et~al.}(2005{\natexlab{a}})}]{Aubert:2005ce}
\bibinfo{author}{\bibfnamefont{B.}~\bibnamefont{Aubert}} \bibnamefont{et~al.}
  (\bibinfo{collaboration}{\babar}), \bibinfo{journal}{Phys. Rev.}
  \textbf{\bibinfo{volume}{D72}}, \bibinfo{pages}{072003}
  (\bibinfo{year}{2005}{\natexlab{a}}), \bibinfo{note}{[Erratum-ibid.\ D{\bf
  74} 099903 (2006)]}.

\bibitem[{\citenamefont{Garmash et~al.}(2006)}]{Garmash:2005rv}
\bibinfo{author}{\bibfnamefont{A.}~\bibnamefont{Garmash}} \bibnamefont{et~al.}
  (\bibinfo{collaboration}{Belle}), \bibinfo{journal}{Phys. Rev. Lett.}
  \textbf{\bibinfo{volume}{96}}, \bibinfo{pages}{251803}
  (\bibinfo{year}{2006}).

\bibitem[{\citenamefont{Garmash et~al.}(2007)}]{Garmash:2006fh}
\bibinfo{author}{\bibfnamefont{A.}~\bibnamefont{Garmash}} \bibnamefont{et~al.}
  (\bibinfo{collaboration}{Belle}), \bibinfo{journal}{Phys. Rev.}
  \textbf{\bibinfo{volume}{D75}}, \bibinfo{pages}{012006}
  (\bibinfo{year}{2007}).

\bibitem[{\citenamefont{Aubert et~al.}(2007{\natexlab{b}})}]{Aubert:2007vi}
\bibinfo{author}{\bibfnamefont{B.}~\bibnamefont{Aubert}} \bibnamefont{et~al.}
  (\bibinfo{collaboration}{\babar}) (\bibinfo{year}{2007}{\natexlab{b}}),
  \eprint{arXiv:0708.2097 [hep-ex]}.

\bibitem[{\citenamefont{Aubert et~al.}(2007{\natexlab{c}})}]{Aubert:2007bs}
\bibinfo{author}{\bibfnamefont{B.}~\bibnamefont{Aubert}} \bibnamefont{et~al.}
  (\bibinfo{collaboration}{\babar}) (\bibinfo{year}{2007}{\natexlab{c}}),
  \eprint{arXiv:0711.4417 [hep-ex]}.

\bibitem[{\citenamefont{Lipkin et~al.}(1991)\citenamefont{Lipkin, Nir, Quinn,
  and Snyder}}]{Lipkin:1991st}
\bibinfo{author}{\bibfnamefont{H.~J.} \bibnamefont{Lipkin}},
  \bibinfo{author}{\bibfnamefont{Y.}~\bibnamefont{Nir}},
  \bibinfo{author}{\bibfnamefont{H.~R.} \bibnamefont{Quinn}}, \bibnamefont{and}
  \bibinfo{author}{\bibfnamefont{A.}~\bibnamefont{Snyder}},
  \bibinfo{journal}{Phys. Rev.} \textbf{\bibinfo{volume}{D44}},
  \bibinfo{pages}{1454} (\bibinfo{year}{1991}).

\bibitem[{\citenamefont{Deshpande et~al.}(2003)\citenamefont{Deshpande, Sinha,
  and Sinha}}]{Deshpande:2002be}
\bibinfo{author}{\bibfnamefont{N.~G.} \bibnamefont{Deshpande}},
  \bibinfo{author}{\bibfnamefont{N.}~\bibnamefont{Sinha}}, \bibnamefont{and}
  \bibinfo{author}{\bibfnamefont{R.}~\bibnamefont{Sinha}},
  \bibinfo{journal}{Phys. Rev. Lett.} \textbf{\bibinfo{volume}{90}},
  \bibinfo{pages}{061802} (\bibinfo{year}{2003}).

\bibitem[{\citenamefont{Ciuchini et~al.}(2006)\citenamefont{Ciuchini, Pierini,
  and Silvestrini}}]{Ciuchini:2006kv}
\bibinfo{author}{\bibfnamefont{M.}~\bibnamefont{Ciuchini}},
  \bibinfo{author}{\bibfnamefont{M.}~\bibnamefont{Pierini}}, \bibnamefont{and}
  \bibinfo{author}{\bibfnamefont{L.}~\bibnamefont{Silvestrini}},
  \bibinfo{journal}{Phys. Rev.} \textbf{\bibinfo{volume}{D74}},
  \bibinfo{pages}{051301} (\bibinfo{year}{2006}).

\bibitem[{\citenamefont{Gronau et~al.}(2007)\citenamefont{Gronau, Pirjol, Soni,
  and Zupan}}]{Gronau:2006qn}
\bibinfo{author}{\bibfnamefont{M.}~\bibnamefont{Gronau}},
  \bibinfo{author}{\bibfnamefont{D.}~\bibnamefont{Pirjol}},
  \bibinfo{author}{\bibfnamefont{A.}~\bibnamefont{Soni}}, \bibnamefont{and}
  \bibinfo{author}{\bibfnamefont{J.}~\bibnamefont{Zupan}},
  \bibinfo{journal}{Phys. Rev.} \textbf{\bibinfo{volume}{D75}},
  \bibinfo{pages}{014002} (\bibinfo{year}{2007}).

\bibitem[{\citenamefont{Gronau et~al.}(2008)\citenamefont{Gronau, Pirjol, Soni,
  and Zupan}}]{Gronau:2007vr}
\bibinfo{author}{\bibfnamefont{M.}~\bibnamefont{Gronau}},
  \bibinfo{author}{\bibfnamefont{D.}~\bibnamefont{Pirjol}},
  \bibinfo{author}{\bibfnamefont{A.}~\bibnamefont{Soni}}, \bibnamefont{and}
  \bibinfo{author}{\bibfnamefont{J.}~\bibnamefont{Zupan}},
  \bibinfo{journal}{Phys. Rev.} \textbf{\bibinfo{volume}{D77}},
  \bibinfo{pages}{057504} (\bibinfo{year}{2008}).

\bibitem[{\citenamefont{Bediaga
  et~al.}(2007{\natexlab{a}})\citenamefont{Bediaga, Guerrer, and
  de~Miranda}}]{Bediaga:2007zz}
\bibinfo{author}{\bibfnamefont{I.}~\bibnamefont{Bediaga}},
  \bibinfo{author}{\bibfnamefont{G.}~\bibnamefont{Guerrer}}, \bibnamefont{and}
  \bibinfo{author}{\bibfnamefont{J.~M.} \bibnamefont{de~Miranda}},
  \bibinfo{journal}{Phys. Rev.} \textbf{\bibinfo{volume}{D76}},
  \bibinfo{pages}{073011} (\bibinfo{year}{2007}{\natexlab{a}}).

\bibitem[{\citenamefont{Garmash et~al.}(2005)}]{Garmash:2004wa}
\bibinfo{author}{\bibfnamefont{A.}~\bibnamefont{Garmash}} \bibnamefont{et~al.}
  (\bibinfo{collaboration}{Belle}), \bibinfo{journal}{Phys. Rev.}
  \textbf{\bibinfo{volume}{D71}}, \bibinfo{pages}{092003}
  (\bibinfo{year}{2005}).

\bibitem[{\citenamefont{Aubert et~al.}(2006)}]{Aubert:2006nu}
\bibinfo{author}{\bibfnamefont{B.}~\bibnamefont{Aubert}} \bibnamefont{et~al.}
  (\bibinfo{collaboration}{\babar}), \bibinfo{journal}{Phys. Rev.}
  \textbf{\bibinfo{volume}{D74}}, \bibinfo{pages}{032003}
  (\bibinfo{year}{2006}).

\bibitem[{\citenamefont{Aubert et~al.}(2007{\natexlab{d}})}]{Aubert:2007sd}
\bibinfo{author}{\bibfnamefont{B.}~\bibnamefont{Aubert}} \bibnamefont{et~al.}
  (\bibinfo{collaboration}{\babar}), \bibinfo{journal}{Phys. Rev. Lett.}
  \textbf{\bibinfo{volume}{99}}, \bibinfo{pages}{161802}
  (\bibinfo{year}{2007}{\natexlab{d}}).

\bibitem[{\citenamefont{Aubert et~al.}(2007{\natexlab{e}})}]{Aubert:2007xb}
\bibinfo{author}{\bibfnamefont{B.}~\bibnamefont{Aubert}} \bibnamefont{et~al.}
  (\bibinfo{collaboration}{\babar}), \bibinfo{journal}{Phys. Rev. Lett.}
  \textbf{\bibinfo{volume}{99}}, \bibinfo{pages}{221801}
  (\bibinfo{year}{2007}{\natexlab{e}}).

\bibitem[{\citenamefont{Minkowski and Ochs}(2005)}]{Minkowski:2004xf}
\bibinfo{author}{\bibfnamefont{P.}~\bibnamefont{Minkowski}} \bibnamefont{and}
  \bibinfo{author}{\bibfnamefont{W.}~\bibnamefont{Ochs}},
  \bibinfo{journal}{Eur. Phys. J.} \textbf{\bibinfo{volume}{C39}},
  \bibinfo{pages}{71} (\bibinfo{year}{2005}).

\bibitem[{\citenamefont{Furman et~al.}(2005)\citenamefont{Furman, Kaminski,
  Lesniak, and Loiseau}}]{Furman:2005xp}
\bibinfo{author}{\bibfnamefont{A.}~\bibnamefont{Furman}},
  \bibinfo{author}{\bibfnamefont{R.}~\bibnamefont{Kaminski}},
  \bibinfo{author}{\bibfnamefont{L.}~\bibnamefont{Lesniak}}, \bibnamefont{and}
  \bibinfo{author}{\bibfnamefont{B.}~\bibnamefont{Loiseau}},
  \bibinfo{journal}{Phys. Lett.} \textbf{\bibinfo{volume}{B622}},
  \bibinfo{pages}{207} (\bibinfo{year}{2005}).

\bibitem[{\citenamefont{Cheng et~al.}(2006)\citenamefont{Cheng, Chua, and
  Yang}}]{Cheng:2005nb}
\bibinfo{author}{\bibfnamefont{H.-Y.} \bibnamefont{Cheng}},
  \bibinfo{author}{\bibfnamefont{C.-K.} \bibnamefont{Chua}}, \bibnamefont{and}
  \bibinfo{author}{\bibfnamefont{K.-C.} \bibnamefont{Yang}},
  \bibinfo{journal}{Phys. Rev.} \textbf{\bibinfo{volume}{D73}},
  \bibinfo{pages}{014017} (\bibinfo{year}{2006}).

\bibitem[{\citenamefont{Wang}(2007)}]{Wang:2006ri}
\bibinfo{author}{\bibfnamefont{Z.-G.} \bibnamefont{Wang}},
  \bibinfo{journal}{Nucl. Phys.} \textbf{\bibinfo{volume}{A791}},
  \bibinfo{pages}{106} (\bibinfo{year}{2007}).

\bibitem[{\citenamefont{Cheng et~al.}(2007)\citenamefont{Cheng, Chua, and
  Soni}}]{Cheng:2007si}
\bibinfo{author}{\bibfnamefont{H.-Y.} \bibnamefont{Cheng}},
  \bibinfo{author}{\bibfnamefont{C.-K.} \bibnamefont{Chua}}, \bibnamefont{and}
  \bibinfo{author}{\bibfnamefont{A.}~\bibnamefont{Soni}},
  \bibinfo{journal}{Phys. Rev.} \textbf{\bibinfo{volume}{D76}},
  \bibinfo{pages}{094006} (\bibinfo{year}{2007}).

\bibitem[{\citenamefont{Klempt and Zaitsev}(2007)}]{Klempt:2007cp}
\bibinfo{author}{\bibfnamefont{E.}~\bibnamefont{Klempt}} \bibnamefont{and}
  \bibinfo{author}{\bibfnamefont{A.}~\bibnamefont{Zaitsev}},
  \bibinfo{journal}{Phys. Rept.} \textbf{\bibinfo{volume}{454}},
  \bibinfo{pages}{1} (\bibinfo{year}{2007}).

\bibitem[{\citenamefont{Aubert et~al.}(2002{\natexlab{a}})}]{babar}
\bibinfo{author}{\bibfnamefont{B.}~\bibnamefont{Aubert}} \bibnamefont{et~al.}
  (\bibinfo{collaboration}{\babar}), \bibinfo{journal}{Nucl. Instrum. Meth.}
  \textbf{\bibinfo{volume}{A479}}, \bibinfo{pages}{1}
  (\bibinfo{year}{2002}{\natexlab{a}}).

\bibitem[{\citenamefont{Kozanecki}(2000)}]{pep2}
\bibinfo{author}{\bibfnamefont{W.}~\bibnamefont{Kozanecki}},
  \bibinfo{journal}{Nucl. Instrum. Meth.} \textbf{\bibinfo{volume}{A446}},
  \bibinfo{pages}{59} (\bibinfo{year}{2000}).

\bibitem[{\citenamefont{Fleming}(1964)}]{isobar1}
\bibinfo{author}{\bibfnamefont{G.~N.} \bibnamefont{Fleming}},
  \bibinfo{journal}{Phys. Rev.} \textbf{\bibinfo{volume}{135}},
  \bibinfo{pages}{B551} (\bibinfo{year}{1964}).

\bibitem[{\citenamefont{Morgan}(1968)}]{isobar2}
\bibinfo{author}{\bibfnamefont{D.}~\bibnamefont{Morgan}},
  \bibinfo{journal}{Phys. Rev.} \textbf{\bibinfo{volume}{166}},
  \bibinfo{pages}{1731} (\bibinfo{year}{1968}).

\bibitem[{\citenamefont{Herndon et~al.}(1975)\citenamefont{Herndon, Soding, and
  Cashmore}}]{isobar3}
\bibinfo{author}{\bibfnamefont{D.}~\bibnamefont{Herndon}},
  \bibinfo{author}{\bibfnamefont{P.}~\bibnamefont{Soding}}, \bibnamefont{and}
  \bibinfo{author}{\bibfnamefont{R.~J.} \bibnamefont{Cashmore}},
  \bibinfo{journal}{Phys. Rev.} \textbf{\bibinfo{volume}{D11}},
  \bibinfo{pages}{3165} (\bibinfo{year}{1975}).

\bibitem[{\citenamefont{Blatt and Weisskopf}(1952)}]{blatt-weisskopf}
\bibinfo{author}{\bibfnamefont{J.}~\bibnamefont{Blatt}} \bibnamefont{and}
  \bibinfo{author}{\bibfnamefont{V.~E.} \bibnamefont{Weisskopf}},
  \emph{\bibinfo{title}{Theoretical Nuclear Physics}} (\bibinfo{publisher}{J.
  Wiley (New York)}, \bibinfo{year}{1952}).

\bibitem[{\citenamefont{Bugg}()}]{buggpc}
\bibinfo{author}{\bibfnamefont{D.~V.} \bibnamefont{Bugg}},
  \bibinfo{note}{private communication.}

\bibitem[{\citenamefont{Flatt\'e}(1976)}]{Flatte}
\bibinfo{author}{\bibfnamefont{S.~M.} \bibnamefont{Flatt\'e}},
  \bibinfo{journal}{Phys. Lett.} \textbf{\bibinfo{volume}{B63}},
  \bibinfo{pages}{224} (\bibinfo{year}{1976}).

\bibitem[{\citenamefont{Ablikim et~al.}(2005)}]{new-BES}
\bibinfo{author}{\bibfnamefont{M.}~\bibnamefont{Ablikim}} \bibnamefont{et~al.}
  (\bibinfo{collaboration}{BES}), \bibinfo{journal}{Phys. Lett.}
  \textbf{\bibinfo{volume}{B607}}, \bibinfo{pages}{243} (\bibinfo{year}{2005}).

\bibitem[{\citenamefont{Aston et~al.}(1988)}]{LASS}
\bibinfo{author}{\bibfnamefont{D.}~\bibnamefont{Aston}} \bibnamefont{et~al.}
  (\bibinfo{collaboration}{LASS}), \bibinfo{journal}{Nucl. Phys.}
  \textbf{\bibinfo{volume}{B296}}, \bibinfo{pages}{493} (\bibinfo{year}{1988}).

\bibitem[{\citenamefont{Bugg}(2003)}]{bugg}
\bibinfo{author}{\bibfnamefont{D.~V.} \bibnamefont{Bugg}},
  \bibinfo{journal}{Phys. Lett.} \textbf{\bibinfo{volume}{B572}},
  \bibinfo{pages}{1} (\bibinfo{year}{2003}).

\bibitem[{\citenamefont{Gounaris and Sakurai}(1968)}]{GS}
\bibinfo{author}{\bibfnamefont{G.~J.} \bibnamefont{Gounaris}} \bibnamefont{and}
  \bibinfo{author}{\bibfnamefont{J.~J.} \bibnamefont{Sakurai}},
  \bibinfo{journal}{Phys. Rev. Lett.} \textbf{\bibinfo{volume}{21}},
  \bibinfo{pages}{244} (\bibinfo{year}{1968}).

\bibitem[{\citenamefont{Zemach}(1964)}]{Zemach1}
\bibinfo{author}{\bibfnamefont{C.}~\bibnamefont{Zemach}},
  \bibinfo{journal}{Phys. Rev.} \textbf{\bibinfo{volume}{133}},
  \bibinfo{pages}{B1201} (\bibinfo{year}{1964}).

\bibitem[{\citenamefont{Zemach}(1965)}]{Zemach2}
\bibinfo{author}{\bibfnamefont{C.}~\bibnamefont{Zemach}},
  \bibinfo{journal}{Phys. Rev.} \textbf{\bibinfo{volume}{140}},
  \bibinfo{pages}{B97} (\bibinfo{year}{1965}).

\bibitem[{\citenamefont{Asner}(2003)}]{asner-review}
\bibinfo{author}{\bibfnamefont{D.}~\bibnamefont{Asner}} (\bibinfo{year}{2003}),
  \eprint{hep-ex/0410014}.

\bibitem[{\citenamefont{Asner et~al.}(2004)}]{Asner:2003uz}
\bibinfo{author}{\bibfnamefont{D.~M.} \bibnamefont{Asner}} \bibnamefont{et~al.}
  (\bibinfo{collaboration}{CLEO}), \bibinfo{journal}{Phys. Rev.}
  \textbf{\bibinfo{volume}{D70}}, \bibinfo{pages}{091101}
  (\bibinfo{year}{2004}).

\bibitem[{\citenamefont{Aubert et~al.}(2002{\natexlab{b}})}]{legendre}
\bibinfo{author}{\bibfnamefont{B.}~\bibnamefont{Aubert}} \bibnamefont{et~al.}
  (\bibinfo{collaboration}{\babar}), \bibinfo{journal}{Phys. Rev. Lett.}
  \textbf{\bibinfo{volume}{89}}, \bibinfo{pages}{281802}
  (\bibinfo{year}{2002}{\natexlab{b}}).

\bibitem[{\citenamefont{Aubert et~al.}(2004{\natexlab{b}})}]{tagging}
\bibinfo{author}{\bibfnamefont{B.}~\bibnamefont{Aubert}} \bibnamefont{et~al.}
  (\bibinfo{collaboration}{\babar}), \bibinfo{journal}{Phys. Rev. Lett.}
  \textbf{\bibinfo{volume}{93}}, \bibinfo{pages}{231801}
  (\bibinfo{year}{2004}{\natexlab{b}}).

\bibitem[{\citenamefont{Aubert et~al.}(2007{\natexlab{f}})}]{Aubert:2007jn}
\bibinfo{author}{\bibfnamefont{B.}~\bibnamefont{Aubert}} \bibnamefont{et~al.}
  (\bibinfo{collaboration}{\babar}), \bibinfo{journal}{Phys. Rev.}
  \textbf{\bibinfo{volume}{D76}}, \bibinfo{pages}{012004}
  (\bibinfo{year}{2007}{\natexlab{f}}).

\bibitem[{\citenamefont{Albrecht et~al.}(1990)}]{argus}
\bibinfo{author}{\bibfnamefont{H.}~\bibnamefont{Albrecht}} \bibnamefont{et~al.}
  (\bibinfo{collaboration}{ARGUS}), \bibinfo{journal}{Z. Phys.}
  \textbf{\bibinfo{volume}{C48}}, \bibinfo{pages}{543} (\bibinfo{year}{1990}).

\bibitem[{\citenamefont{Abe et~al.}(2002)}]{Abe:2002av}
\bibinfo{author}{\bibfnamefont{K.}~\bibnamefont{Abe}} \bibnamefont{et~al.}
  (\bibinfo{collaboration}{Belle}), \bibinfo{journal}{Phys. Rev.}
  \textbf{\bibinfo{volume}{D65}}, \bibinfo{pages}{092005}
  (\bibinfo{year}{2002}).

\bibitem[{\citenamefont{Aubert et~al.}(2004{\natexlab{c}})}]{Aubert:2003mi}
\bibinfo{author}{\bibfnamefont{B.}~\bibnamefont{Aubert}} \bibnamefont{et~al.}
  (\bibinfo{collaboration}{\babar}), \bibinfo{journal}{Phys. Rev.}
  \textbf{\bibinfo{volume}{D70}}, \bibinfo{pages}{092001}
  (\bibinfo{year}{2004}{\natexlab{c}}).

\bibitem[{\citenamefont{Aubert et~al.}(2007{\natexlab{g}})}]{Aubert:2007si}
\bibinfo{author}{\bibfnamefont{B.}~\bibnamefont{Aubert}} \bibnamefont{et~al.}
  (\bibinfo{collaboration}{\babar}), \bibinfo{journal}{Phys. Rev.}
  \textbf{\bibinfo{volume}{D76}}, \bibinfo{pages}{031103}
  (\bibinfo{year}{2007}{\natexlab{g}}).

\bibitem[{\citenamefont{Jen et~al.}(2006)}]{Jen:2006in}
\bibinfo{author}{\bibfnamefont{C.~M.} \bibnamefont{Jen}} \bibnamefont{et~al.}
  (\bibinfo{collaboration}{Belle}), \bibinfo{journal}{Phys. Rev.}
  \textbf{\bibinfo{volume}{D74}}, \bibinfo{pages}{111101}
  (\bibinfo{year}{2006}).

\bibitem[{\citenamefont{Akhmetshin et~al.}(2007)}]{Akhmetshin:2006bx}
\bibinfo{author}{\bibfnamefont{R.~R.} \bibnamefont{Akhmetshin}}
  \bibnamefont{et~al.} (\bibinfo{collaboration}{CMD-2}),
  \bibinfo{journal}{Phys. Lett.} \textbf{\bibinfo{volume}{B648}},
  \bibinfo{pages}{28} (\bibinfo{year}{2007}).

\bibitem[{\citenamefont{Bediaga
  et~al.}(2007{\natexlab{b}})\citenamefont{Bediaga, Boito, Guerrer, Navarra,
  and Nielsen}}]{Bediaga:2007mi}
\bibinfo{author}{\bibfnamefont{I.}~\bibnamefont{Bediaga}},
  \bibinfo{author}{\bibfnamefont{D.~R.} \bibnamefont{Boito}},
  \bibinfo{author}{\bibfnamefont{G.}~\bibnamefont{Guerrer}},
  \bibinfo{author}{\bibfnamefont{F.~S.} \bibnamefont{Navarra}},
  \bibnamefont{and} \bibinfo{author}{\bibfnamefont{M.}~\bibnamefont{Nielsen}}
  (\bibinfo{year}{2007}{\natexlab{b}}), \eprint{arXiv:0709.0075 [hep-ph]}.

\bibitem[{\citenamefont{Pivk and Le~Diberder}(2005)}]{Pivk:2004ty}
\bibinfo{author}{\bibfnamefont{M.}~\bibnamefont{Pivk}} \bibnamefont{and}
  \bibinfo{author}{\bibfnamefont{F.~R.} \bibnamefont{Le~Diberder}},
  \bibinfo{journal}{Nucl. Instrum. Meth.} \textbf{\bibinfo{volume}{A555}},
  \bibinfo{pages}{356} (\bibinfo{year}{2005}).

\bibitem[{\citenamefont{Aubert et~al.}(2005{\natexlab{b}})}]{Aubert:2005cp}
\bibinfo{author}{\bibfnamefont{B.}~\bibnamefont{Aubert}} \bibnamefont{et~al.}
  (\bibinfo{collaboration}{\babar}), \bibinfo{journal}{Phys. Rev.}
  \textbf{\bibinfo{volume}{D71}}, \bibinfo{pages}{111101}
  (\bibinfo{year}{2005}{\natexlab{b}}).

\bibitem[{\citenamefont{Aubert et~al.}(2007{\natexlab{h}})}]{Aubert:2007mj}
\bibinfo{author}{\bibfnamefont{B.}~\bibnamefont{Aubert}} \bibnamefont{et~al.}
  (\bibinfo{collaboration}{\babar}), \bibinfo{journal}{Phys. Rev. Lett.}
  \textbf{\bibinfo{volume}{99}}, \bibinfo{pages}{021603}
  (\bibinfo{year}{2007}{\natexlab{h}}).

\end{thebibliography}
\bibliographystyle{apsrev}

\end{document}